\documentclass{aa}  

\usepackage{graphicx}
\usepackage{subcaption}
\usepackage{amsmath}
\usepackage{siunitx}
\usepackage{bm}
\usepackage[backref=page]{hyperref} 
\hypersetup{breaklinks=true}

\usepackage{txfonts}

\usepackage{multirow}
\usepackage{diagbox} 
\usepackage{adjustbox}
\usepackage{float}
\usepackage{array}  
\usepackage{tabularx} 
\usepackage{booktabs}   
\usepackage{placeins}
\usepackage{pdflscape}

\usepackage{xcolor}

\usepackage{setspace} 

\usepackage{natbib,twoopt}

\bibpunct{(}{)}{;}{a}{}{,}
\makeatletter
\newcommandtwoopt{\citeads}[3][][]{\href{http://adsabs.harvard.edu/abs/#3}%
{\def\hyper@linkstart##1##2{}%
\let\hyper@linkend\@empty\citealp[#1][#2]{#3}}}
\newcommandtwoopt{\citepads}[3][][]{\href{http://adsabs.harvard.edu/abs/#3}%
{\def\hyper@linkstart##1##2{}%
\let\hyper@linkend\@empty\citep[#1][#2]{#3}}}
\newcommandtwoopt{\citetads}[3][][]{\href{http://adsabs.harvard.edu/abs/#3}%
{\def\hyper@linkstart##1##2{}%
\let\hyper@linkend\@empty\citet[#1][#2]{#3}}}
\newcommandtwoopt{\citeyearads}[3][][]%
{\href{http://adsabs.harvard.edu/abs/#3}
{\def\hyper@linkstart##1##2{}%
\let\hyper@linkend\@empty\citeyear[#1][#2]{#3}}}
\makeatother

\begin{document} 

  \title{Bulk viscous cosmological models with a cosmological constant: Observational constraints}
  \titlerunning{Bulk viscous cosmological models with $\Lambda$}

   \author{R. Noemí Villalobos
          \inst{1}
          \and
          Yerko Vásquez\inst{2}
          \and
          Norman Cruz \inst{3,4}
          \and
           Carlos H. López-Caraballo \inst{5,6}
          }
   \authorrunning{Villalobos et al.}
   \institute{Departamento de Astronomía, Universidad de La Serena, Raúl Bitrán 1305, La Serena, Chile\\
              \email{rvillalobosr@userena.cl}
         \and 
              Departamento de Física, Universidad de La Serena, Avenida Juan Cisternas 1200, La Serena, Chile\\
              \email{yvasquez@userena.cl}
         \and
             Departamento de Física, Universidad de Santiago de Chile, Avenida Víctor Jara 3493, Estación Central, 9170124 Santiago, Chile\\
             \email{norman.cruz@usach.cl}
        \and 
             Center for Interdisciplinary Research in Astrophysics and Space Exploration (CIRAS), Universidad de Santiago de Chile, Avenida Libertador Bernardo O’Higgins 3363, Estación Central, Chile
        \and
            Instituto de Astrofísica de Canarias, E-38200 La Laguna, Tenerife, Spain
        \and
            Departamento de Astrofísica, Universidad de La Laguna, E-38206 La Laguna, Tenerife, Spain\\
             }

   \date{Received ...mes dia..., 2025; accepted ...mes dia, 2025}

  \abstract{}
    {We investigate whether viscous cold dark matter (vCDM) in a $\Lambda$-dominated Friedmann-Lemaître-Robertson-Walker (FLRW) universe can alleviate the Hubble tension, while satisfying thermodynamic constraints. Here, we examine both flat and curved geometries.}
    {We modeled a vCDM cosmology with a bulk viscosity $\zeta = \zeta_0\,(\Omega_{\rm{vc}}/\Omega_{\rm{vc}0})^m$, where $m$ determines the viscosity evolution and $\Omega_{vc}$ is the density parameter of vCDM. We explored two particular scenarios: (i) constant viscosity $\zeta = \zeta_0$ ($m=0$) and (ii) variable viscosity $\zeta = \zeta(\Omega_{\rm{vc}})$ ($m$ free).  Using Bayesian inference, we constrained these models with the latest datasets: the Pantheon+ Type Ia supernova sample (with a SH0ES calibration, PPS, and without, PP), Hubble parameter measurements, $H(z)$, from cosmic chronometers and baryon acoustic oscillations (BAO) as independent datasets, including DESI DR2, and a Gaussian prior on 
    $H_0$ from the SH0ES measurement, $H_0=73.04 \pm 1.04$ ~\si{km.s^{-1}.Mpc^{-1}} (R22 prior).  We compared the models via Akaike, Bayesian, and deviance  information criteria (AIC, BIC, and DIC), and with Bayesian evidence.} 
    {Our results indicate that the Hubble tension persists, although it shows partial alleviation ($\sim 1\sigma$ tension) in all investigated scenarios when local measurements are included. For the flat $m=0$ case, the joint analysis yields $H_0 = 71.05^{+0.62}_{-0.60}$~km~s$^{-1}$~Mpc$^{-1}$. Curved models initially favor $\Omega_{\rm{K0}} > 0$ (at more than $2\sigma$), but this preference shifts toward flatness once the PPS + R22 prior is included. Notably, the current viscosity is constrained to $\zeta_0 \sim 10^6$~\si{Pa.s} in all scenarios, in agreement with the thermodynamic requirements. Although the model selection via BIC and Bayesian evidence favors $\Lambda$CDM, the AIC and DIC show mild support for viscous models in some datasets.}
   {Bulk viscous models moderately improve the fits, but they can neither resolve the Hubble tension nor outperform the $\Lambda$CDM model.
    To achieve more robust constraints, future analyses should incorporate CMB observations, which are expected to break parameter degeneracies involving $m$ and $\tilde{\zeta}_0$.}
  {}

   \maketitle
   \nolinenumbers

\section{Introduction}

\defcitealias{2022ApJ...934L...7R}{R22}

The most widely accepted cosmological model that remains consistent with observational data is the lambda-cold dark matter ($\Lambda$CDM) model of a spatially flat universe. This model proposes that the cosmological constant $\Lambda$, assumed to represent dark energy, drives the accelerated expansion of the Universe \citep{1992ARA&A..30..499C, 1998AJ....116.1009R, 1999ApJ...517..565P, 2001LRR.....4....1C, 2003RvMP...75..559P, 2008ARA&A..46..385F, 2013PhR...530...87W, 2020A&A...641A...6P}, while cold dark matter (CDM) behaves as a nonrelativistic, pressureless perfect fluid. However, the $\Lambda$CDM  model faces several significant issues \citep{1989RvMP...61....1W}. For example, it cannot explain the physical nature of CDM or $\Lambda$ \citep{1989RvMP...61....1W, 2003PhR...380..235P} and, additionally, it fails to solve the fine-tuning problem; specifically, the question of why $\Lambda$ is so small compared to the vacuum energy predicted by quantum field theory \citep{1968SvPhU..11..381Z, 1989RvMP...61....1W, 2003PhR...380..235P, 2021PhRvD.104l3517S}. It also does not address the coincidence problem, which asks why $\Lambda$ does not become dynamically relevant until the late Universe \citep{1999PhRvL..82..896Z}. Collectively, these unresolved issues are known as the cosmological constant problem \citep{1989RvMP...61....1W}.

With the increasing precision of cosmological observations, several tensions related to key parameters have emerged; most notably the so-called  "Hubble tension", which arises from the inconsistency between the values of the Hubble constant $H_0$ derived from early Universe measurements and local (late-time) observations. The Planck Collaboration, using observations of the cosmic microwave background (CMB) within the $\Lambda$CDM framework, reported $H_0 = (67.36 \pm 0.54)$~\si{km.s^{-1}.Mpc^{-1}} at a $68\%$ confidence level (CL).  In contrast, the SH0ES Collaboration (Supernovae $H_0$ for the equation of state, EoS) found a significantly higher value of $H_0 = (73.04 \pm 1.04)$~\si{km.s^{-1}.Mpc^{-1}} at $68\%$ CL, using the three-rung distance ladder method with Cepheids \citep{2022ApJ...934L...7R}. These results reveal a tension of approximately $5\sigma$, which is unlikely to be due to systematic errors and, thus, strongly suggests there is a need for new physics beyond the $\Lambda$CDM model. A comprehensive review of proposals aimed at addressing the Hubble tension can be found in \citep{2021CQGra..38o3001D, 2022LRR....25....6M, 2024ARA&A..62..287V, 2024PhRvD.110l3518P, 2022ApJ...933..212M}.

For these reasons, in recent decades, a wide variety of alternatives to the standard cosmological model have been proposed. In most of these proposals, cosmic components are modeled as perfect fluids. Among them, scalar field models with a dynamical EoS have attracted significant attention.  Quintessence models offer a dynamical alternative to $\Lambda$ (e.g., \citep{1998PhRvL..80.1582C, 2019PDU....2600385D}), while phantom \citep{2002PhLB..545...23C} and k-essence models \citep{2002PhRvD..66f3514C} extend the framework to EoS parameters $\omega < -1$ or to noncanonical kinetic terms. Additionally, various reviews of dark energy and modified gravity models can be found in \citep{2006IJMPD..15.1753C, 2012Ap&SS.342..155B}. 

Among the alternative approaches, a less conventional one modifies the dark sector by introducing dissipative terms into the energy-momentum tensor to generate accelerated expansion \citep{2010JCAP...08..009A, 2017PhRvD..96l4020C, 2023PDU....4201351C}, while also attempting to alleviate current cosmological tensions \citep{2017JCAP...11..005A, 2019PhRvD.100j3518Y, 2023ApJ...959..120A}. These contributions include heat flux \citep{2019CQGra..36u5002B}, bulk viscous pressure, and shear viscosity. The most extensively studied cosmological models with viscous fluids incorporate bulk viscosity, as they remain compatible with the cosmological principle. In contrast, shear viscosity introduces tensor terms that explicitly violate isotropy.

Among the theories that describe viscous fluids, the first to address bulk viscosity was Eckart's theory \citep{1940PhRv...58..919E}. Although this formalism suffers from well-known causality issues, its mathematical simplicity has made it widely studied, providing a fundamental first-order approximation for dissipative fluids.  Israel and Stewart later proposed a causal second-order theory \citep{ISRAEL1976310, ISRAEL1979341}, introducing relaxation time as a new parameter. 

A bulk viscous fluid is characterized by two key thermodynamic quantities: the energy density, $\rho$, and the effective pressure, $P_{\rm{eff}} = p + \Pi$, where $p$ is the equilibrium pressure and $\Pi$ denotes the bulk viscous pressure. Within Eckart's first-order formalism, the viscous pressure takes the form $\Pi = -\zeta \vartheta$, where $\zeta$ is the bulk viscosity coefficient and $\vartheta \equiv u^{\mu}_{;\mu} = 3H$ is the expansion scalar, with $H \equiv \dot{a}/a$ being the Hubble parameter. The Israel-Stewart theory extends this description by introducing a propagation equation for $\Pi$ that resolves the causality limitations of Eckart's approach. 

Both the Eckart and Israel-Stewart theories can account for the accelerated expansion of the Universe without a cosmological constant, provided that $\Pi < 0$ (and, consequently, $\zeta > 0$) to satisfy the second law of thermodynamics \citep{1972gcpa.book.....W}. However, analyses of dissipative inflation \citep{1995CQGra..12.1455M} reveal a fundamental limitation: models relying exclusively on bulk viscosity to drive acceleration violate the near-equilibrium condition $|\Pi/p| \ll 1$, which is required to maintain the system close to local thermal equilibrium. This violation implies that purely viscous mechanisms cannot realistically drive cosmic acceleration without additional components.

This motivates our inclusion of $\Lambda$, supported by observational evidence for its cosmological role  \citep{1998AJ....116.1009R, 1999ApJ...517..565P, 2013ApJS..208...20B}. Recent studies \citep{2018JCAP...12..017C, 2020EPJC...80..637H, 2022PhRvD.105b4047C, 2022Symm...14.1866C} show that a positive $\Lambda$ can preserve near-equilibrium conditions in certain regimes, although this requires moving beyond unified viscous dark matter models as complete descriptions of late-time cosmology.

Given that cosmic matter exhibits dissipation \citep{2018JCAP...05..031A},  here we analyze viscous cold dark matter (vCDM) models within Eckart's theory in the presence of $\Lambda$, following the background dynamics framework presented in \citep{2012PhRvD..86h3501V, 2014PhRvD..90l3526V}. We denote these models as $\Lambda$vCDM.  We explore two particular scenarios: (i) constant bulk viscosity ($\zeta = \zeta_0$, where $\zeta_0$ is the present-day value) and (ii) variable  bulk viscosity ($\zeta = \zeta(\Omega_{\rm{vc}}))$, where $\Omega_{\rm{vc}}$ denotes the vCDM energy density parameter). We study both scenarios in flat and curved FLRW universes, where $\Pi$ modifies the effective energy-momentum tensor. Therefore, this framework extends the $\Lambda$CDM model by introducing dissipative effects in the dark matter component.

This paper is organized as follows. In Section ~\ref{sec:theory}, we explain the theoretical framework underpinning bulk viscous models. We describe the methodology and datasets employed to constrain the cosmological parameters in Section~\ref{sec:methodology-dataset}. In Section.~\ref{sec:results}, we present and discuss the resulting constraints on various cosmological
parameters, including their concordance with the benchmark $\Lambda$CDM model, evaluated using information criteria and Bayesian inference.  Finally, in Section~\ref{sec:conclusions}, we summarize our conclusions.

\section{Background for bulk viscous models}
\label{sec:theory}

A homogeneous and isotropic universe is described by the FLRW metric,  given by
 
\begin{equation}
    ds^{2} = - c^{2}dt^{2} + a^{2}(t) \left(\frac{dr^{2}}{1 - Kr^{2}} + r^{2} d\theta^{2} + r^{2} \sin^{2}\theta d\phi^{2} \,\right)\,,
\label{eq:FLRW_metric}
\end{equation}

\noindent
where $(r, \theta, \phi)$ are the comoving coordinates, $a(t)$ is the scale factor at cosmic time, $t$, and $K$ denotes the spatial curvature parameter. The values of $K = -1, 0, +1$ correspond to an open, flat, and closed universe, respectively. \\

We assume a universe containing a component that experiences dissipative processes exclusively through bulk viscosity. This introduces a nonadiabatic contribution ($\Delta T^{\mu\nu}$) to the perfect fluid energy-momentum tensor \citep{1972gcpa.book.....W}. Therefore, the total energy-momentum tensor for a bulk viscous fluid takes the form of

\begin{equation}
\begin{aligned}
    T^{\mu\nu} &\equiv \mathcal{T}^{\mu\nu} +  \Delta T^{\mu\nu} \\
    &= \left(\rho + P_{\rm {eff}}\right) u^{\mu}u^{\nu} + P_{\rm {eff}}\,g^{\mu\nu},
\end{aligned}
\label{eq:tensor_bulk}
\end{equation}

\noindent
where $\mathcal{T}^{\mu\nu}$ is the energy-momentum tensor that describes a perfect fluid and the four-velocity  satisfies the normalization condition $u^{\mu} u_{\mu} = -1$. In the Eckart framework, bulk viscosity modifies the equilibrium pressure $p$ to an effective pressure,

\begin{equation}
    P_{\rm {eff}} =  p - 3H\zeta,
\end{equation}

\noindent
where $p \equiv  \omega\rho$ follows a barotropic EoS. The dimensionless parameter $\omega$ can be constant (e.g., $\omega = 0$ for CDM or $\omega = -1$ for $\Lambda$) or time-dependent, as in dynamical dark energy models \citep[see][]{1998PhRvL..80.1582C, 2002PhRvD..66f3514C}. 

When vCDM is the only viscous component in a universe with noninteracting cosmic fluids, its energy-momentum conservation (derived from $\nabla_{\mu}T^{\mu\nu} = 0$) yields

\begin{equation}
    \dot{\rho}_{\rm{vc}} + 3H\rho_{\rm{vc}} =  9\zeta H^2 \,,
    \label{eq:rho_vc}
\end{equation}

\noindent
where $\rho_{\rm{vc}}$ and $p_{\rm{vc}} = 0$ are the energy density and pressure of the vCDM component (abbreviated as `vc' in equations), respectively.  The corresponding  Friedmann equations, including spatial curvature and the cosmological constant, are 

\begin{align}
\label{eq:Hubble_parameter}
    & H^2 = \dfrac{8\pi G}{3c^2}\,\left(\rho_{\rm{r}} + \rho_{\rm{b}} + \rho_{\rm{vc}}\right) - \dfrac{Kc^2}{a^2} + \dfrac{\Lambda c^2}{3},\\
    & 2\dot{H} + 3H^2 =   - \dfrac{8\pi G}{c^2} \left(p_{\rm{r}} + p_{\rm{b}} + p_{\rm{vc}} -  3\zeta H\right) - \dfrac{Kc^2}{a^2} + \Lambda c^2\,, 
\label{eq:aceleration_equation}
\end{align}

\noindent
where $G$ is Newton's gravitational constant. The subscripts $r$ and $b$ denote radiation and baryonic matter, respectively. Combining the scale factor-redshift relation $a(t) = (1+z)^{-1}$ with Eq. (\ref{eq:Hubble_parameter}) yields the normalized Hubble parameter $E(z) \equiv H(z)/H_0$ for our $\Lambda$vCDM model, including spatial curvature, expressed as

\begin{equation}
\begin{aligned}
    E(z) = \Big[\Omega_{\rm{r0}}\,(1+z)^4  + \Omega_{\rm{vc}}(z) &+ \Omega_{\rm{b0}} (1+z)^3  \\
    &  + \Omega_{\rm{K0}} (1+z)^2 + \Omega_{\Lambda}\Big]^{1/2}\,,
\end{aligned}
\label{eq:E(z)}
\end{equation}

\noindent
where $H_0 = 100\,h$~\si{km.s^{-1}.Mpc^{-1}} is the Hubble constant, while $h$ is the dimensionless Hubble parameter. The energy density parameter of vCDM is defined as

\begin{equation}
    \Omega_{\rm{vc}}(z) \equiv \dfrac{\rho_{\rm{vc}}}{\rho_{\rm{crit},0}}= \dfrac{8\pi G\, \rho_{\rm{vc}}(z)}{3H_0^2\,c^2}\,,
    \label{eq:Omega_vc}
\end{equation}

\noindent
and its present-day is $\Omega_{\rm{vc}}(z=0) = \Omega_{\rm{vc}0}$. The remaining set, $\Omega_{\rm{r0}}$, $\Omega_{\rm{b0}}$, $\Omega_{\rm{K}0}$, and $\Omega_{\Lambda} \equiv \Omega_{\rm{de}}(z=0)$ are defined as
\begin{equation}
    \begin{aligned}
       &\Omega_{\rm{r0}} = \dfrac{8\pi G \rho_{\rm{r0}}}{3H_0^2 c^2}; \quad \Omega_{\rm{b0}}= \dfrac{8\pi G  \rho_{\rm{b0}}}{3H_0^2 c^2};\\
       &\Omega_{\rm{K0}} = \dfrac{8\pi G  \rho_{\rm{K}0}}{3H_0^2 c^2}; \quad \Omega_{\Lambda} = \dfrac{8\pi G  \rho_{\Lambda}}{3H_0^2 c^2}\,. 
    \end{aligned}
\end{equation}

\noindent
The sum of these present-day density parameters satisfies the constraint,  
\begin{equation}
   \Omega_{\rm{r0}} + \Omega_{\rm{b0}} + \Omega_{\rm{vc}0}  + \Omega_{\rm{K0}} + \Omega_{\Lambda} = 1\,. 
\end{equation} 

For radiation, we use $\Omega_{\rm{r0}} \approx \left(1 + 0.2271~N_{\rm{eff}}\right)\Omega_{\gamma 0}$, where $\Omega_{\gamma 0} = 2.469 \cdot 10^{-5} ~ h^{-2}$ \citep{2011ApJS..192...18K} is the present-day photon density parameter and $N_{\rm{eff}} = 3.046$  is the effective number of neutrino species according to the standard $\Lambda$CDM model \citep{2002PhLB..534....8M, 2005NuPhB.729..221M}. 

We can rewrite Eq. (\ref{eq:rho_vc}) in terms of dimensionless quantities, as in \citep{2012PhRvD..86h3501V, 2014PhRvD..90l3526V}, and solve it numerically by imposing the initial condition, $\Omega_{\rm{vc}}(z = 0) = \Omega_{\rm{vc}0}$. Introducing the notation ${}' \equiv d/dz$, we obtain

\begin{equation}
\begin{aligned}
    (1+z)\,\Omega_{\rm{vc}}' & - 3 \Omega_{\rm{vc}}(z) + \zeta^{\ast}\,\left[\Omega_{\rm{r0}}\,(1+z)^4 + \Omega_{\rm{vc}}(z) \right.\\ 
    &\left. + \Omega_{\rm{b0}}(1+z)^3 + \Omega_{\rm{K}0}(1+z)^2 + \Omega_{\Lambda}\right]^{1/2} = 0\,,
\end{aligned}
\label{eq:conser_equation_vc}
\end{equation}

\noindent
where we have defined the dimensionless  viscosity parameter,

\begin{equation}
    \zeta^{\ast} = \dfrac{24 \pi \, G }{c^2}\,\dfrac{\zeta}{H_0} \,.
    \label{eq:zeta_ast}
\end{equation}

\noindent
We also derive the corresponding evolution equation for the Hubble parameter,

\begin{equation}
\begin{aligned}
    2(1+z)&H H'\,-\, 3H^2  + \zeta^{\ast} H\,H_0\\ 
    &-  H_0^2  \,\Big[\Omega_{\rm{r0}}\,(1+z)^4
    -\,\Omega_{\rm{K0}}\,(1+z)^2  \,-\, 3\Omega_{\Lambda}\Big]= 0\,.
\end{aligned}
\label{eq:evol_Hubble}
\end{equation}

Each model is characterized by the bulk viscosity parameter $\zeta$, as defined in Eq. (\ref{eq:zeta_ast}), and it is reduced to the standard $\Lambda$CDM model when $\zeta = 0$. We then adopt a power law dependence of viscosity on vCDM density,

\begin{equation}
    \zeta  = \zeta_0 \left(\dfrac{\Omega_{\rm{vc}}}{\Omega_{\rm{vc}0}}\right)^m\,,
    \label{eq:ansatz_zeta}
\end{equation}

\noindent
where $m$ is the viscous exponent and $\zeta_0$ denotes the present-day bulk viscosity coefficient. Substituting the viscosity ansatz Eq. (\ref{eq:ansatz_zeta}) into Eq. (\ref{eq:zeta_ast}), we obtain \citep{2012PhRvD..86h3501V}:

\begin{equation}
    \zeta^{\ast} = \tilde{\zeta_{0}} \,\left(\dfrac{\Omega_{\rm{vc}}}{\Omega_{\rm{vc0}}}\right)^m, \quad \text{with} \quad \tilde{\zeta}_0 = \dfrac{24 \pi \, G }{c^2}\,\dfrac{\zeta_0}{H_0}\,.
    \label{eq:zeta0_tilde}
\end{equation}

\noindent
Here, $\tilde{\zeta}_0$ is the dimensionless bulk viscosity parameter. In SI units, $\zeta$ has dimensions of Pascal-seconds (\si{Pa.s}), while $H_0$ is expressed in inverse seconds (\si{{s}^{-1}}).
 
In this work, we study bulk viscous models with $\Lambda$, considering both spatially flat and curved geometries, labeled as $\Lambda$vCDM and $\Lambda$vCDM + $\Omega_{\rm{K}}$, respectively. Based on Eq. (\ref{eq:zeta0_tilde}), we analyze two cases: (i) a constant bulk viscosity with $m=0$, representing the simplest scenario; and (ii) a variable bulk viscosity with $m$  treated as a free parameter. This defines four distinct bulk viscous models: $\Lambda$vCDM ($m=0$), $\Lambda$vCDM + $\Omega_{\rm{K}}$ ($m=0$), $\Lambda$vCDM ($m$ free), and $\Lambda$vCDM + $\Omega_{\rm{K}}$ ($m$ free). Additionally, we analyze the $\tilde{\zeta}_0 = 0$ case (the benchmark $\Lambda$CDM model) in Appendix \ref{Appendix:LCDM_model}. The free parameters for each model  are listed in Table~\ref{table:free_parameters-priors}.

\begin{table*}[h!]
\caption{Free parameters and priors used in the MCMC analysis.}
\label{table:free_parameters-priors}
\centering
\begin{tabular}{llll}
\hline \midrule
    \multicolumn{2}{l}{Model} & Parameter space & Prior\\  \midrule
    \multirow{2}{*}{$m=0$} & $\Lambda$vCDM &  $\Theta = \{H_0, \Omega_{\rm{b0}}h^2, \Omega_{\rm{vc0}}h^2, M, \tilde{\zeta}_{0}\}$  & $\pi(\Omega_{\rm{vc}0}h^2)$; $\pi(\Omega_{\rm{dm}0}h^2)$:  $\mathcal{N}$(0.12, 0.01)  \\ 
    \noalign{\smallskip}
    & $\Lambda\rm{vCDM} + \Omega_{\rm{K}}$  &  $\Theta = \{H_0,  \Omega_{\rm{K0}}, \Omega_{\rm{b0}}h^2, \Omega_{\rm{vc0}}h^2,  M, \tilde{\zeta}_{0}\}$ &   $\pi(\Omega_{\rm{b0}}h^2)$: $\mathcal{N}$(0.022, 0.01)\\ \cmidrule{1-3}
    \multirow{2}{*}{$m$ free} &  $\Lambda$vCDM & $\Theta = \{H_0,  \Omega_{\rm{b0}}h^2, \Omega_{\rm{vc0}}h^2, M, m, \tilde{\zeta}_{0}\}$ & $\pi(\tilde{\zeta}_{0})$: $\mathcal{U}(0, 1)$ \\  
    \noalign{\smallskip}
    & $\Lambda\rm{vCDM} + \Omega_{\rm{K}}$ &  $\Theta = \{H_0, \Omega_{\rm{K0}},  \Omega_{\rm{b0}}h^2, \Omega_{\rm{vc0}}h^2, M, m, \tilde{\zeta}_{0}\}$ & $\pi(m)$: $\mathcal{U}(-2, 2)$ \\ \cmidrule{1-3}
     \multirow{2}{*}{$\tilde{\zeta}_0 =0$} & $\Lambda$CDM & $\Theta = \{H_0,  \Omega_{\rm{b0}}h^2, \Omega_{\rm{dm0}}h^2,  M\}$ &  $\pi(H_0)$: $\mathcal{U}(30, 99)$   \\
     \noalign{\smallskip}
     & $\Lambda$CDM + $\Omega_{\rm{K}}$ & $\Theta = \{H_0, \Omega_{\rm{K0}},  \Omega_{\rm{b0}}h^2, \Omega_{\rm{dm0}}h^2, M\}$ &  $\pi(\Omega_{\rm{K}0})$: $\mathcal{U}(-0.4, 0.4)$ ; \ $\pi(M)$: $\mathcal{U}(-30,-5)$ \\
\midrule
\end{tabular}
\tablefoot{ We list the prior ranges, using $\mathcal{N}(\mu, \sigma)$ to denote a Gaussian and $\mathcal{U}(a,b)$ to denote a uniform prior}
\end{table*}

\section{Methodology and dataset for observational constraints}
\label{sec:methodology-dataset}

In this section, we focus on constraining the bulk viscous models using publicly available observational data. To achieve this, we utilized recent datasets, including Hubble parameter measurements from cosmic chronometers (CC) and baryon acoustic oscillations (BAOs), along with  Type Ia supernovae (SNe Ia) data. The subsections below describe each dataset and the applied methodology.

\subsection{Bayesian statistical procedure}
\label{sec:statistics}

Bayesian inference is a robust statistical technique for parameter estimation and model comparison, widely employed in cosmology. Based on Bayes' theorem, the posterior probability of a model $\mathcal{M}$ with free parameters, $\Theta$, given a dataset, $\mathcal{D}$, is expressed as

\begin{equation}
    \mathcal{P}(\Theta\,|\,\mathcal{D}, \mathcal{M}) = \dfrac{\mathcal{L}(\mathcal{D}\,|\,\Theta, \mathcal{M})\,\pi(\Theta\,|\,\mathcal{M})}{\mathcal{E}(\mathcal{D}\,|\,\mathcal{M})}\,,
\end{equation}

\noindent
where $\mathcal{P}(\Theta\,|\,\mathcal{D}, \mathcal{M})$ is the posterior distribution, $\mathcal{L}(\mathcal{D} | \Theta,\mathcal{M})$ is the likelihood, $\pi(\Theta | \mathcal{M})$ is the prior, and $\mathcal{E}(\mathcal{D}|\mathcal{M})$ is the evidence (also known as Bayesian evidence, marginal likelihood, or model likelihood). 

A  Gaussian likelihood  is related to  $\chi^2$ statistic (usually called chi-square) via the relation $\mathcal{L} \propto e^{-\chi^2/2}$,  where $\chi^2$  quantifies the discrepancy between observed data and model predictions. The best-fit parameters, corresponding to the maximum likelihood estimate, can be obtained by minimizing the $\chi^2$ function.

\subsubsection{Bayesian evidence}

The evidence is a normalization constant in parameter estimation, which is crucial for model comparison.  It is obtained by integrating the likelihood multiplied by the prior over the full parameter space,

\begin{equation}
    \mathcal{E}(\mathcal{D}|\mathcal{M}) = \int \,d\Theta \,\mathcal{L}(\mathcal{D}\,|\,\Theta, \mathcal{M})\, \pi(\Theta\,|\,\mathcal{M}).
\end{equation}

For two competing models, $\mathcal{M}_0$ and $\mathcal{M}_i$, the relative posterior probabilities (or posterior odds) determine which model is favored by the data, $\mathcal{D}$,

\begin{equation}
    \dfrac{\mathcal{P}(\mathcal{M}_0\,|\,\mathcal{D})}{\mathcal{P}(\mathcal{M}_i\,|\,\mathcal{D})} = \dfrac{\mathcal{E}(\mathcal{D}\,|\,\mathcal{M}_0)\,\pi(\mathcal{M}_0)}{\mathcal{E}(\mathcal{D}\,|\,\mathcal{M}_i)\,\pi(\mathcal{M}_i)} = \mathcal{B}_{0i} \,\dfrac{\pi(\mathcal{M}_0)}{\pi(\mathcal{M}_i)} ,
\end{equation}

\noindent
where $\mathcal{B}_{0i} \equiv \mathcal{E}_0/\mathcal{E}_i$  is the Bayes factor, which quantifies the relative evidence for the  models. Values of  $\mathcal{B}_{0i} > 1$ indicate an increase in the support for $\mathcal{M}_0$ over $\mathcal{M}_i$, while $\mathcal{B}_{0i} < 1$ favors $\mathcal{M}_i$.

We adopted the Jeffreys scale, as suggested by \citep{2008ConPh..49...71T, 2017PhRvL.119j1301H}, to interpret the strength of evidence, as summarized in Table~\ref{table:strength_evidence}. When $\ln \mathcal{B}_{0i} > 1$,
this suggests a preference for $\mathcal{M}_0$; conversely, when $\ln \mathcal{B}_{0i} < -1$, the data support the alternative model, $\mathcal{M}_i$.

\begin{table}[t!]
\caption{Jeffreys scale for the Bayes factor ($\mathcal{B}_{0i}$).}
\label{table:strength_evidence}
    \centering
    \begin{tabular}{ccl}
    \midrule\midrule
        $|\ln \mathcal{B}_{0i}|$ & Probability & Strength of evidence\\
    \midrule
        $< 1.0$ & $< 0.750$  & Inconclusive\\
         $1.0 - 2.5$ & 0.750  & Weak evidence \\
        $2.5 - 5.0$ & $0.923$  & Moderate evidence\\
        $ \geq 5.0$ & $0.993$  & Strong evidence \\
    \midrule
    \end{tabular}
\tablefoot{We use this scale used to quantify the strength of evidence when comparing the reference model $\mathcal{M}_{0}$ against an alternative model $\mathcal{M}_{i}$. The right column lists the thresholds and their associated evidence levels.}
\end{table}

\subsubsection{Information criteria for the model comparison}

A natural question is how efficient these models are compared to the standard $\Lambda$CDM model. To address this,  we employed the Akaike information criterion (AIC; \citealt{1974ITAC...19..716A}), Bayesian information criterion (BIC; \citealt{1978AnSta...6..461S}), and the deviance information criterion (DIC; \citealt{RePEc:bla:jorssb:v:64:y:2002:i:4:p:583-639}). The AIC and BIC quantify the trade-off between goodness of fit and model complexity by penalizing the inclusion of additional free parameters. They are defined, respectively, as

\begin{align}
\label{eq:AIC}
    \mathrm{AIC} &= \chi^2_{\rm{min}} + 2k,\\
    \mathrm{BIC} &= \chi^2_{\rm{min}} + k \ln N,
\label{eq:BIC}
\end{align}

\noindent
where $\chi^2_{\rm{min}}$ is the minimum chi-square value obtained from the best-fit parameters, $k$ represents the number of free parameters, and $N$ is the total number of data points used in the fit. The BIC penalizes additional parameters more strongly than the AIC, thus favoring simpler models. In both cases, the preferred model is the one that minimizes the corresponding information criterion.

The DIC combines concepts from Bayesian statistics and information theory. Unlike the AIC and BIC, the DIC determines the complexity penalty from the posterior distribution. It is defined as

\begin{equation}
    \rm{DIC} \equiv D(\overline{\Theta}) + 2\mathcal{C}_B = \overline{D(\Theta)} + \mathcal{C}_B,
    \label{eq:DIC}
\end{equation}

\noindent
where $\mathcal{C}_B = \overline{D(\Theta)} - D(\overline{\Theta})$ is the Bayesian complexity, quantifying the effective number of parameters. The overlines denote posterior mean values, and $D(\Theta)$ is the Bayesian deviance.  For Gaussian likelihoods, the deviance is reduced to $D(\Theta) = \chi^2(\Theta)$.

We can evaluate the relative strength of evidence for each candidate model $i$ by calculating the differences in information criteria ($\Delta$IC) relative to the benchmark $\Lambda$CDM model, defined as

\begin{equation}
    \Delta \rm{IC} = \rm{IC}_{\Lambda\rm{CDM}} - \rm{IC}_{i}.
    \label{eq:Delta_IC}
\end{equation}

\noindent
With this definition, positive values of $\Delta$IC indicate a preference  for model $i$ over $\Lambda$CDM.  We interpret the strength of evidence according to the standard  scale summarized in Table~\ref{table:information_criteria}.  

\begin{table}[t!]
\caption{Interpretation based on the $\Delta \rm{IC}$ criteria.}
\label{table:information_criteria}
    \centering
    \begin{tabular}{cl}
    \midrule \midrule
    $\Delta\rm{AIC}$ / $\Delta\rm{BIC}$ / $\Delta\rm{DIC}$ & Strength of evidence\\ \midrule
        $0 - 2$ & Weak  \\
        $2 - 6$ & Positive \\
        $6 - 10$ & Strong \\
        $> 10$ & Very strong \\
    \midrule
    \end{tabular}
\tablefoot{A value of $\Delta\rm{IC}>~2$  indicates preference for the bulk viscous model, $i$, against $\Lambda$CDM.}
\end{table}

\subsection{Observational datasets}

\subsubsection{Cosmic chronometers}

The most direct way to constrain the cosmological parameters is using H(z) data.  We compiled a sample of 33 data points from various sources (shown in Table \ref{table:CC_data}), measured in the redshift range $0.07 \leq z \leq 1.965$, where the total error per redshift is given by $\sigma_{\rm{tot}} = [\sigma_{\rm{stat}}^2 + \sigma_{\rm{syst}}^2]^{1/2}$. These H(z) data points were obtained using the CC method, which relies on comparing the differential age evolution of galaxies at different redshifts, as proposed by \citep{2002ApJ...573...37J}. The method estimates the Hubble rate based on the relationship,

\begin{equation}
    H(z) = - \dfrac{1}{1+z} \dfrac{dz}{dt}\,.
    \label{CC_method}
\end{equation}

This approach is independent of cosmological assumptions, making it a valuable tool for testing different models by directly measuring the expansion rate of the Universe.

By constructing the chi-square function, we effectively constrain the free parameters of the bulk viscous models. To perform the MCMC analysis, the $\chi^2$ function for the CC method is given by

\begin{equation}
    \chi_{\rm{CC}}^2 (\Theta)= \sum_{i = 1}^{33} \left(\frac{ H^{\rm{th}}(z_i, \Theta) - H^{\rm{obs}}_{\rm{CC}}(z_i)}{\sigma_{i, \rm{CC}}}\right)^2  \,,
\end{equation}

\noindent
where $H_{\rm{th}}(z_i, \Theta)$ represents the theoretical value of the Hubble parameter at redshift, $z_i$, with the free model parameters, $\Theta$,  shown in Table \ref{table:free_parameters-priors}, $H^{\rm{obs}}_{\rm{CC}}(z_i)$ denotes the corresponding observed Hubble data values at $z_i$, and $\sigma_{i, \rm{CC}}$ represents the observational uncertainty associated with $H^{\rm{obs}}_{\rm{CC}}(z_i)$.

\subsubsection{Hubble constant}

We included the SH0ES Collaboration value of the Hubble constant, $H_0 = (73.04 \pm 1.04)$~\si{km.s^{-1}.Mpc^{-1}} (hereafter R22), from \cite{2022ApJ...934L...7R} as a Gaussian prior in our analysis.

\subsubsection{Type Ia Supernovae}

SNe Ia exhibit a well-known brightness, making them ideal standard candles for measuring cosmological distances. The theoretical distance modulus, $\mu$, at redshift $z$ is given by \citep{1972gcpa.book.....W} and expressed as

\begin{equation} 
    \mu(z)  \equiv m - M =  5 \log_{10} \left[ \frac{d_{\rm{L}}(z)}{1 \rm{Mpc}} \right] +  25 \,,
    \label{eq:dist_modulus_model}
\end{equation}

\noindent
where $m$ is the observed apparent magnitude, $M$ is the absolute magnitude, and $d_L\equiv (c/H_0) D_L$ is the luminosity distance in megaparsecs (Mpc). The luminosity distance is related to the angular diameter distance, $d_A$, via the Etherington distance-duality relation $d_L(z) = (1+z)^2\,d_A$. Both distances can be expressed in terms of the comoving angular diameter distance $D_M(z)$ as 

\begin{equation}
    d_{\rm{L}}(z) = (1+z)\,D_{\rm{M}}(z)\quad \text{and} \quad
    d_{\rm{A}}(z) = \dfrac{D_{\rm{M}}(z)}{(1+z)} \,.
    \label{eq:DL_DA}
\end{equation}

\noindent
Here, 

\begin{equation}
    D_{\rm{M}}(z) \equiv \dfrac{c}{H_0}\,S_{\rm{k}}\left(x\right), \quad \text{with} \quad x = \int_0^z \frac{dz'}{E(z')} \,, 
    \label{eq:DM}
\end{equation}

\noindent
where 
\begin{equation}
S_{\rm{k}}(x) = 
\begin{cases}
\dfrac{1}{\sqrt{\Omega_{\rm{K}0}}} \,\sinh\left( \sqrt{\Omega_{\rm{K}0}}\,x\right), &\text{if} \quad \Omega_{\rm{K}0} > 0,\\
x, &\text{if} \quad \Omega_{\rm{K}0} = 0,\\
\dfrac{1}{\sqrt{|\Omega_{\rm{K}0}|}} \,\sin\left(\sqrt{|\Omega_{\rm{K}0}|} \,x\right), &\text{if} \quad \Omega_{\rm{K}0} < 0.
\end{cases}
\end{equation}

In this study, we used both the Pantheon+ sample (PP) \citep{2022ApJ...938..113S, 2022ApJ...938..110B} and Pantheon+ $\&$SH0ES sample (PPS) \citep{2022ApJ...938..110B}. The PP sample comprises 1701 light curves from 1550 distinct SNe Ia measurements, within the redshift range $0.001 < z < 2.26$, providing apparent magnitudes rather than distance moduli.  We can compute the $\chi^2$ function as

\begin{equation}
\chi^2_{\rm{PP}} = \Delta \vec{D} ^T \cdot C^{-1} \cdot \Delta \vec{D} \,,
\end{equation}

\noindent
where $\vec{D}$ is the residual vector of dimension 1701, with components $\Delta D_i = \mu_i - \mu_{\rm{model}}(z_i, \Theta)$, quantifying the difference between the observed and predicted distance moduli. The total covariance matrix, $C = C_{\rm{stat}}+C_{\rm{syst}}$,  combines the diagonal covariance matrix of the statistical uncertainties $C_{\rm{stat}}$ for $\mu_i(z_i)$ with the covariance matrix of systematic uncertainties, $C_{\rm{syst}}$, for each SN Ia light curve.

The PPS sample combines the original PP datasets with the Cepheid host distance measurements from the SH0ES Collaboration \citep{2022ApJ...934L...7R}. This joint likelihood analysis helps to constrain the degeneracy between the parameters $M$ of SN Ia and $H_0$. 

The inclusion of Cepheid host distances modifies the SN distance residuals as 
\begin{equation}
\Delta D'_i = 
\begin{cases}
 \mu_i- \mu_i^{\rm{Cepheid}}, & i \,\in\, \text{Cepheid hosts},\\
 \mu_i - \mu_{\rm{model}}(z_i, \Theta), & \text{otherwise},
\end{cases}
\end{equation}

\noindent
where $\mu_i^{\rm{Cepheid}}$ denotes the Cepheid-calibrated distance modulus to the host galaxies. This modification leads to the following $\chi^2$ function,

\begin{equation}
\chi^2_{\rm{PPS}} = \Delta \vec{D}'^T \cdot (C^{\,\rm{SN Ia}}\,+\, C^{\rm{\,Cepheid}})^{-1}\cdot \Delta \vec{D}' \,.
\end{equation}

\noindent
where $C^{\rm{Cepheid}} = C_{\rm{stat}}^{\rm{Cepheid}} +C_{\rm{syst}}^{\rm{Cepheid}}$ is the SH0ES Cepheid host-distance covariance matrix. 

The SNe Ia sample includes both heliocentric and CMB-corrected redshifts. Therefore, the luminosity distance is computed by incorporating these redshift corrections as

\begin{equation}
    d_{\rm{L}} (z, \Theta) = (1 + z_{\rm{CMB}})(1 + z_{\rm{hel}})\,d_{\rm{A}}(z, \Theta) \,.
\end{equation}

\noindent
Taking into account possible contributions from spatial curvature, we can numerically compute $d_A(z, \Theta)$ for each $z_i$ and parameter set $\Theta$ by solving the differential equation,

\begin{equation}
\begin{aligned}
    d_{\rm{A}}'(z, \Theta)  =  -\,&\dfrac{d_{\rm{A}}(z, \Theta)}{(1 + z)}
    \,+\, \dfrac{c}{(1 + z)\,H(z, \Theta)} \cdot\\
    &\cdot \left[ 1 \,+\, \Omega_{\rm{K}0}\,(1 + z)^2\, \left(\, \dfrac{H_0\,d_{\rm{A}}(z, \Theta)}{c}\right)^2\right]^{1/2}\,,
    \label{dif_DA}
\end{aligned}
\end{equation}

\noindent
which follows an analogous formulation to that presented in \citet{2024PDU....4601650F} for luminosity distance calculations.

\subsubsection{Baryonic acoustic oscillations}

Overall, BAOs provide a standard ruler originating from acoustic waves in the photon-baryon fluid of the early Universe. Their characteristic scale corresponds to the sound horizon at the baryon drag epoch, $r_{\rm{d}} \equiv r_{\rm{s}}(z_{\rm{d}})$  \citep{1998ApJ...496..605E}, which is constrained by CMB measurements. This scale is calculated as 

\begin{equation}
   r_{\rm{d}}= \dfrac{1}{H_0}\,\int^{\infty}_{z_{\rm{d}}} \dfrac{c_{\rm{s}}(z)}{E(z)}\,dz,
    \label{Eq:sound_horizon_drag_epoch}
\end{equation}

\noindent
where $c_{\rm{s}}(z)$ is the speed of sound in the photon-baryon fluid. In the $\Lambda$CDM framework, this epoch occurs at $z_{\rm{d}} \approx 1060$ (\citealt{2020A&A...641A...6P}). 

BAO measurements can be separated into directions transverse and radial to the line of sight. This leads to two primary observables, both normalized by the drag scale $r_{\rm{d}}$:
the transverse comoving distance, $D_{\rm{M}}(z)/r_{\rm{d}}$, and the radial Hubble distance, $D_{\rm{H}}(z)/r_{\rm{d}}$, where $D_{\rm{H}}(z) \equiv c/H(z)$. These two measurements are often combined into the spherically averaged distance, $D_V(z) \equiv [z \,D_{\rm{M}}^2(z) \,D_{\rm{H}}(z)]^{1/3}$.

For our analysis, we used BAO measurements from the Dark Energy Spectroscopic Instrument Data Release 2 (DESI DR2), which cover the redshift range $0.1 < z < 4.2$ using multiple tracers, as detailed in Table III of \citep{2025PhRvD.112h3515A}. The dataset includes the bright galaxy sample (BGS; $0.1 < z < 0.4$), luminous red galaxies (LRGs; $0.4 < z < 1.1$), emission-line galaxies (ELGs; $1.1 < z < 1.6$), quasars (QSO; $0.8 < z < 2.1$), and the Lyman-$\alpha$ forest (Ly$\alpha$; $1.77 < z < 4.16$). We refer to this entire dataset as “DESI”. Our sample consists of 13 data points (listed in  Table~\ref{Appendix:DESI_data}). The $\chi^2$ function for the DESI data is then constructed as
\begin{equation}
    \chi^2_{\mathrm{DESI}} = \Delta \vec{Q}^T C_{\rm{DESI}}^{-1} \Delta\vec{Q},
\end{equation}

\noindent
where $C_{\rm{DESI}}$ is the covariance matrix corresponding to the DESI data and $\Delta\vec{Q}$ is the corresponding residual vector. Its components are given by

\begin{equation}
\Delta Q_i = \left(\frac{D_{\alpha} (z_i)}{r_{\rm{d}}}\right)^{\rm{obs}} - \left(\frac{D_{\alpha}(z_i, \Theta)}{r_{\rm{d}}}\right)^{\rm{th}}, \quad \alpha \in \{V, M, H\}.
\end{equation}

Additionally, we considered adopting direct constraints on the expansion history using Hubble parameter measurements derived from BAO. These were obtained by combining the radial BAO scale with an appropriate value of $r_{\rm{d}}$. We used an independent compilation of 30 uncorrelated BAO-based $H(z)$ measurements from various galaxy surveys (listed in Table \ref{table:HBAO_data}), covering $0.24~\leq~z~\leq~2.36$. The $\chi^2$ function for this BAO $H(z)$ dataset is defined as

\begin{equation}
    \chi_{\rm{HBAO}}^2 = \sum_{i = 1}^{30} \left(\frac{ H^{\rm{th}}(z_i, \Theta) - H^{\rm{obs}}_{\rm{HBAO}}(z_i)}{\sigma_{i, \mathrm{HBAO}}}\right)^2,
\end{equation}

\noindent
where $H^{\rm{th}}(z_i, \Theta)$ is the theoretical value of the Hubble parameter at redshift $z_i$, for the free model parameters $\Theta$ (see Table~\ref{table:free_parameters-priors}), $H^{\rm{obs}}_{\rm{HBAO}}(z_i)$ represents the observed Hubble parameter values derived from BAO measurements, and $\sigma_{i, \mathrm{HBAO}}$ denotes the associated uncertainties.

\subsubsection{ Joint analysis}

For joint observational constraints using the R22 prior, CC, BAO, and SNe~Ia datasets, the total $\chi^2_{\rm{JOIN}}$ is given by

\begin{equation}
    \chi^2_{\rm{JOIN}} = \chi_{{\rm{BAO}}}^2 + \chi_{\rm{SNIa}}^2 + \chi_{\rm{CC}}^2 + \chi_{\rm{R22}}^2. 
\end{equation}

\noindent
where $\chi_{\rm{SNIa}}^2$ represents either $\chi_{\rm{PP}}^2$ or $\chi_{\rm{PPS}}^2$, $\chi_{\rm{BAO}}^2$ denotes either $\chi_{\rm{HBAO}}^2$ or their sum $\chi_{\rm{HBAO}}^2 + \chi_{\rm{DESI}}^2$ depending on the dataset combination. We assume that measurements from different databases are uncorrelated.

\section{Results and analysis}
\label{sec:results}

\begin{figure}[t!]
    \centering
    \includegraphics[width=0.45\textwidth]{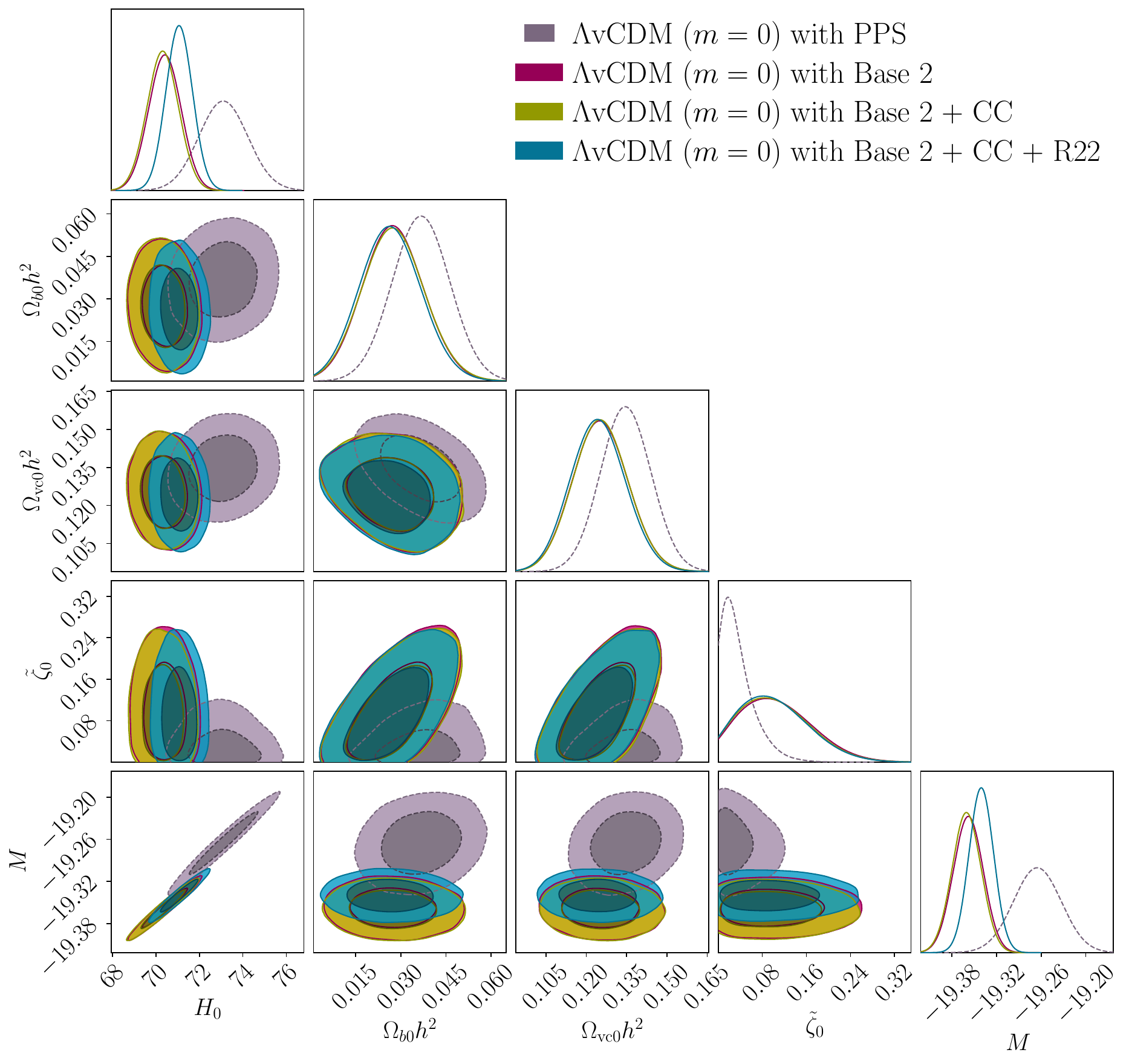}
    \includegraphics[width=0.4\textwidth]{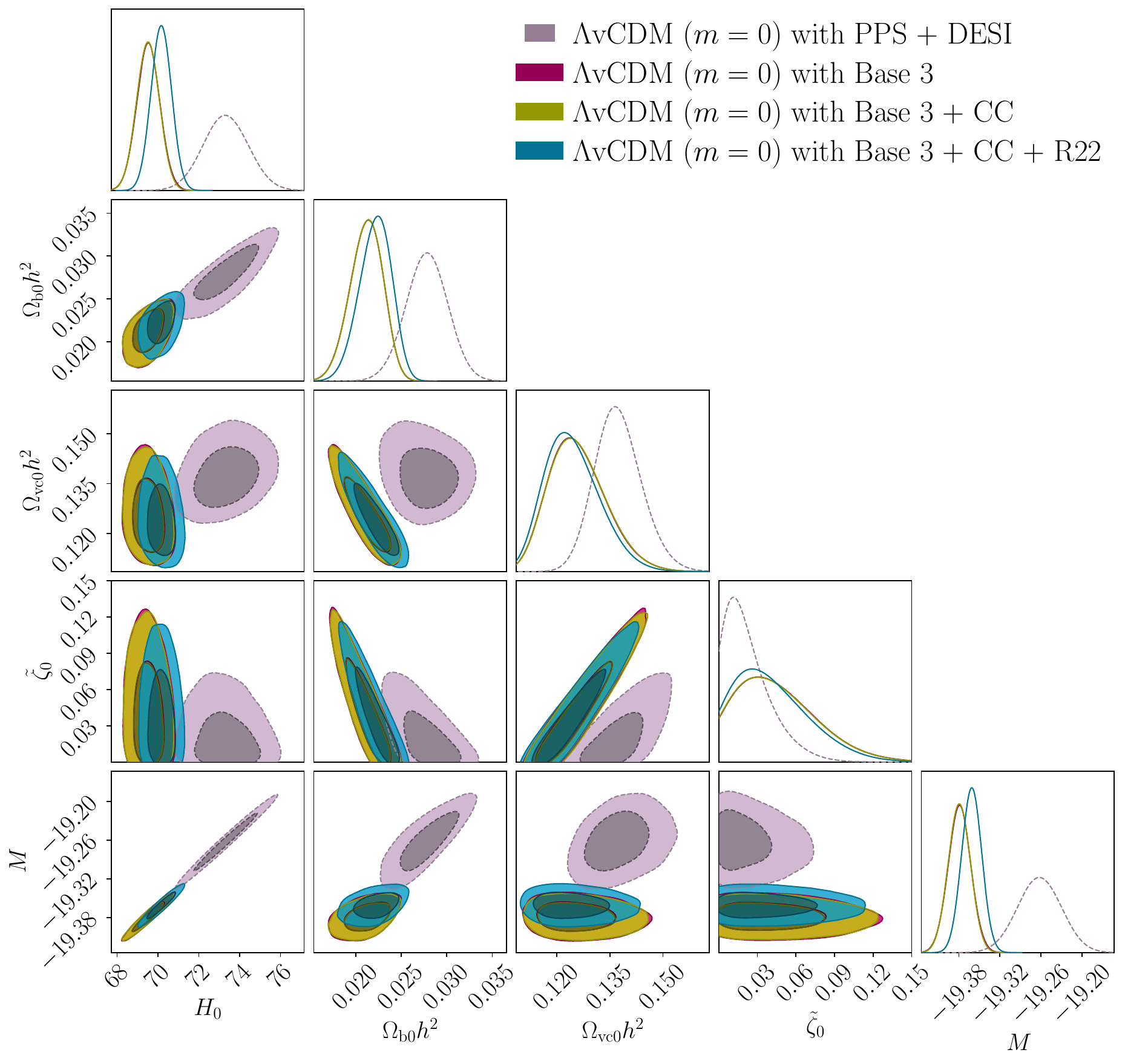}
    \caption{Cosmological parameter constraints for a flat viscous universe with a constant viscosity ($m = 0$). Top: PPS, Base~2, CC, and R22 datasets. Bottom: PPS + DESI, Base 3, CC, and R22 datasets. Contours show 2D confidence regions ($68\%$ and $95\%$ CL)  and their 1D marginalized posterior distributions.}
    \label{fig:flat-LvCDM_m=0_JOIN_BASE1-2}
\end{figure}

In this section, we present the constraints on all the bulk viscous models with a cosmological constant. To minimize the $\chi^2$ function for each model, we performed a Bayesian analysis using the affine-invariant MCMC sampler implemented in the \textsc{emcee} \footnote{\url{https://emcee.readthedocs.io/en/stable/}} Python module \citep{2013ascl.soft03002F}. Our MCMC configuration uses 60 walkers and 30\,000 steps per walker. To ensure convergence, we discarded an initial burn-in period corresponding to ten times the maximum autocorrelation time across all parameters. For the likelihood computation, we incorporated the DESI, PP, and PPS datasets from the public version of \texttt{CosmoSIS} \footnote{\url{https://cosmosis.readthedocs.io/en/latest/}}  \citep{2015A&C....12...45Z}. We derived parameter constraints and generated confidence contours using the ChainConsumer package \footnote{\url{https://samreay.github.io/ChainConsumer/}} \citep{Hinton2016}. 
In this work, we adopted two types of priors in our MCMC analysis: Gaussian priors $\mathcal{N}(\mu, \sigma)$, described by a normal distribution with a mean, $\mu$, and a standard deviation, $\sigma$, and  uniform priors $\mathcal{U}(a, b)$, which follow a top-hat function over the interval $[a, b]$. The specific priors for each free parameter are summarized in Table \ref{table:free_parameters-priors}.

We present the observational analysis using several cosmological datasets to constrain the model parameters and assess their consistency with current observations. Our dataset includes CC, BAO measurements (HBAO and DESI), SNe Ia (PP and PPS), and the R22  prior. We defined three primary dataset combinations: Base 1 (HBAO + PP), Base~2 (HBAO + PPS), and Base~3 (HBAO + PPS + DESI), whose joint chi-square functions are $\chi^2_{\rm{Base1}} = \chi^2_{\rm{HBAO}} + \chi^2_{\rm{PP}}$, $\chi^2_{\rm{Base2}}= \chi^2_{\rm{HBAO}} + \chi^2_{\rm{PPS}}$, and $\chi^2_{\rm{Base3}}= \chi^2_{\rm{HBAO}} + \chi^2_{\rm{PPS}} + \chi^2_{\rm{DESI}}$, respectively. We then extended these combinations by including CC and the R22 prior, evaluating the joint constraints for Base~1 + CC + R22, Base~2 + CC + R22, and Base~3 + CC + R22. 

Figures \ref{fig:flat-LvCDM_m=0_JOIN_BASE1-2}--\ref{fig:LvCDM+OK_mfree_JOIN_BASE1-2} show the contour plots from our MCMC analysis, including both the two-dimensional (2D) confidence regions at the $68\%$ and $95\%$ CL and the 1D marginalized posterior distributions for the free parameters of the models (see Table \ref{table:free_parameters-priors}). Different dataset combinations are represented by different colors: PPS (purple-gray), PPS + DESI (purple), Base~1 (dark magenta), Base~1 + CC (green), and Base~1 + CC + R22 (sea blue). The corresponding Base~2 and Base~3 combinations follow the same color scheme. In Appendix \ref{appendix:summary_constraints}, Tables \ref{table:results_m=0} and \ref{table:results_mfree} summarize the best-fit values and $68\%$ CL uncertainties ($1\sigma$) for the free parameters of both the $\Lambda$vCDM model and its extension $\Lambda$vCDM + $\Omega_{\rm{K}}$.

\subsection{$m=0$ scenario}

\begin{figure}[h!]
\centering
        \includegraphics[width=0.45\textwidth]{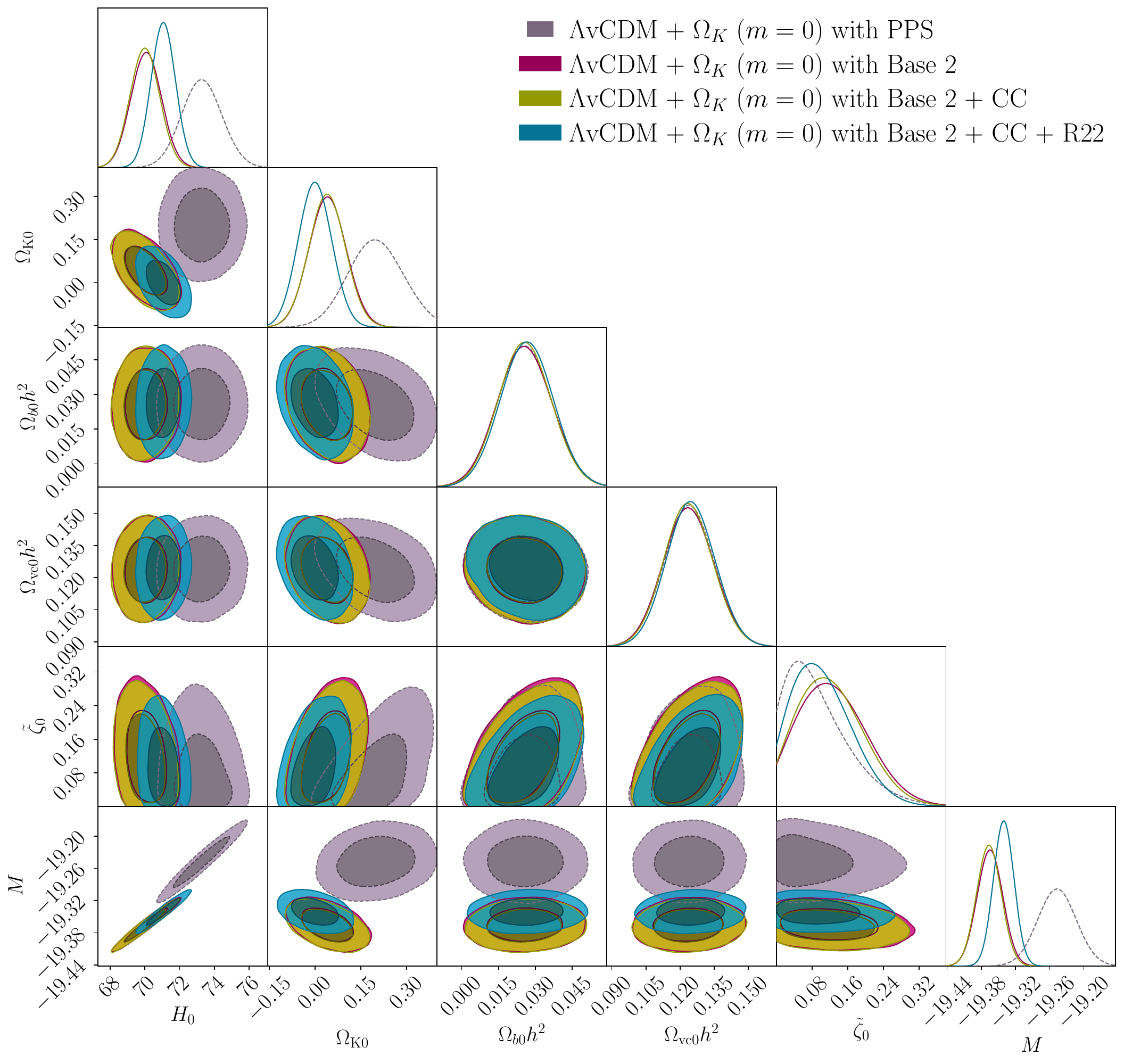}
        \includegraphics[width=0.45\textwidth]{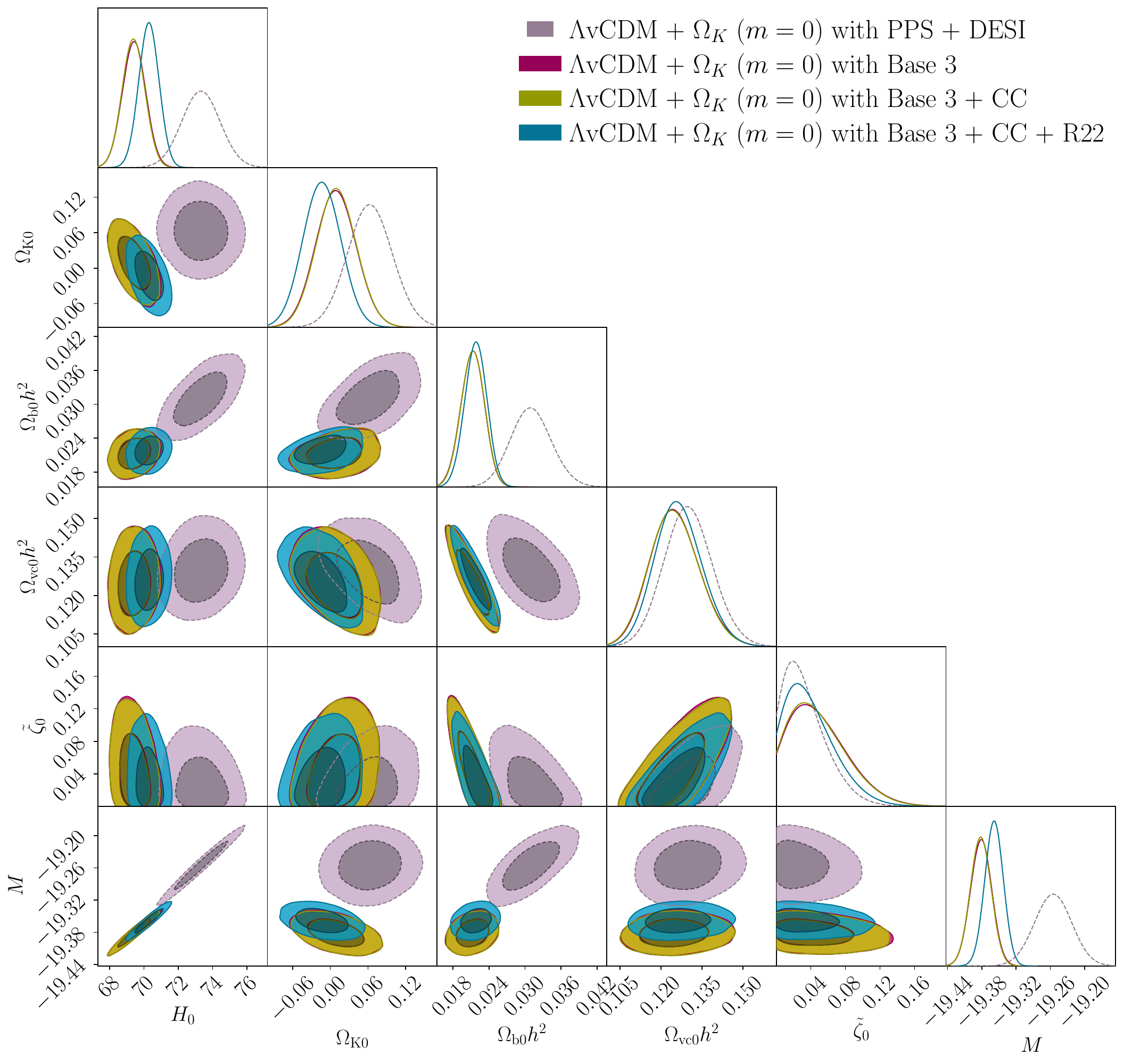}
    \caption{Cosmological parameter constraints for a non-flat viscous universe with constant viscosity ($m = 0$). Top panel: PPS, Base~2, CC, and R22 datasets. Bottom panel: PPS + DESI, Base~3, CC, and R22 datasets. The contours show 2D confidence regions ($68\%$ and $95\%$ CL) and their 1D marginalized posterior distributions.}
    \label{fig:LvCDM+OK_m=0_JOIN_BASE1-2}
\end{figure}

We first analyzed the simplest viscous scenario with $m=0$, corresponding to a constant bulk viscosity. The resulting constraints are summarized in Table~\ref{table:results_m=0}.  Both the flat universe (Fig.~\ref{fig:flat-LvCDM_m=0_JOIN_BASE1-2}) and the curved universe (Fig.~\ref{fig:LvCDM+OK_m=0_JOIN_BASE1-2}) cases exhibit strong positive correlations between $H_0$ and $M$ ($\rm{corr} > 0.95$) across all dataset combinations, as well as a variable correlation with $\Omega_{\rm{K0}}$: negligible for PPS alone ($\rm{corr} = 0.009$), but moderately negative when including additional constraints ($\rm{corr}\sim -0.5 $). In curved models, $H_0$ exhibits a weaker negative correlation with $\tilde{\zeta}_0$, while $\tilde{\zeta}_0$ itself correlates positively with $\Omega_{\rm{K}}$ (e.g., $\rm{corr} = 0.33$ for Base 2 + CC + R22, $\rm{corr} = 0.51$ for PPS, and $\rm{corr} = 0.27$ for PPS + DESI).  

Our analysis of the flat universe scenario in Table \ref{table:results_m=0} shows that Base~1 yields $H_0 = 67.94^{+0.97}_{-0.92}$~\si{km.s^{-1}.Mpc^{-1}}, while Base~2 produces a higher estimate of $H_0$ ($=70.39^{+0.75}_{-0.73}$~\si{km.s^{-1}.Mpc^{-1}}), resulting in a tension of $\sim 2\sigma$. Notably, Base~3 provides an intermediate constraint of $H_0 = 69.53 \pm 0.56$~\si{km.s^{-1}.Mpc^{-1}},  which reduces the tension with Base~1 to $\sim 1.4\sigma$. This trend continues when the CC data is considered. Incorporating the R22 prior reduces these discrepancies to $1.19 \sigma$ between the Base~1 + CC + R22 and Base~2 + CC + R22 combinations, and to just $0.29 \sigma$ between the Base~1 + CC + R22 and Base~3 + CC + R22 datasets, yielding

\begin{subequations}
\begin{align}
    H_0 &= 69.92^{+0.71}_{-0.74}~\si{km.s^{-1}.Mpc^{-1}} \quad (\mathrm{Base~1 + CC + R22}),\\
    H_0 &= 71.05^{+0.62}_{-0.60}~\si{km.s^{-1}.Mpc^{-1}} \quad (\mathrm{Base~2 + CC + R22}),\\
    H_0 & = 70.17^{+0.49}_{-0.51}~\si{km.s^{-1}.Mpc^{-1}} \quad {(\mathrm{Base~3 + CC + R22})},
\end{align}
\end{subequations}

\noindent
suggesting that both the inclusion of local R22 measurements and DESI data help alleviate the tension between the datasets. In the non-flat scenario, we observe systematically lower estimates of $H_0$, specifically $H_0 = 66.1^{+1.2}_{-1.1}$~\si{km.s^{-1}.Mpc^{-1}} for Base~1 and $H_0 = (66.3 \pm 1.1)$~\si{km.s^{-1}.Mpc^{-1}} for Base~1 + CC. However, when we include R22 prior to the Base~2 + CC and Base~3 + CC datasets, the estimates increase to $H_0 = (71.06 \pm 0.70)$~\si{km.s^{-1}.Mpc^{-1}} and $H_0 = (70.28^{+0.60}_{-0.58})$~\si{km.s^{-1}.Mpc^{-1}}, respectively. These values align with those in the flat geometry scenario, with differences smaller than $0.2\sigma$.

We see that the datasets significantly influence the inferred spatial curvature. Incorporating the R22 prior favors a flat universe, yielding

\begin{equation}
    \Omega_{\rm{K}0} = 0.049^{+0.059}_{-0.058} \quad (\mathrm{Base~1+CC+R22}),
\end{equation}

\noindent
which is compatible at $1\sigma$ with a flat universe. Including the PPS sample further sharpens these constraints, yielding

\begin{equation}
    \Omega_{\rm{K0}} = -0.002^{+0.056}_{-0.053} \quad (\mathrm{Base~2+CC+R22}),
\end{equation}

\noindent
indicating only a $0.04\sigma$ deviation from flatness. We further confirm this trend by considering the DESI data, which yield the following results, all fully consistent with a flat universe:

\begin{subequations}
\begin{align}
    \Omega_{\rm{K0}} &= 0.009\pm 0.032 \quad (\mathrm{Base 3}),\\[2pt]
    \Omega_{\rm{K0}} &= 0.009^{+0.032}_{-0.031} \quad (\mathrm{Base 3 + CC}),\\[2pt]
    \Omega_{\rm{K0}} &= -0.014^{+0.030}_{-0.029} \quad (\mathrm{Base 3 + CC + R22}).
\end{align}
\end{subequations}

\noindent
Even with the inclusion of the R22 prior and DESI data, we observe no significant deviation of $\sim 0.5 \sigma$. In contrast, some datasets without DESI favor  $\Omega_{\rm{K}0} > 0$ at more than $2\sigma$, suggesting a moderate preference toward an open universe. For example, the PPS sample alone gives $\Omega_{\rm{K0}} = 0.198^{+0.092}_{-0.092}$ ($2.2\sigma$), while Base~1 yields $\Omega_{\rm{K0}} = 0.181^{+0.073}_{-0.074}$ ($2.5\sigma$). This tension suggests that local measurements impose stricter constraints on spatial flatness, effectively breaking the geometric degeneracy.

In addition, the dimensionless viscosity coefficient $\tilde{\zeta}_0$ shows good agreement with observational data (see Figs. \ref{fig:plots_OHD-evol_H(z)_models} and \ref{fig:plots_data-Dist_modulo_models}). For both the flat and non-flat scenarios, the present-day bulk viscosity $\zeta_0$, computed via Eq. (\ref{eq:zeta0_tilde}), is approximately

\begin{equation}
    \zeta_0 \sim 10^6~\si{Pa.s}, 
\end{equation}

\noindent
in line with estimates from previous studies \citep{ 2017MPLA...3250026N, 2022Symm...14.1866C}. 
This value is consistent with studies that consider only the isotropic and homogeneous background, which allows for relatively large bulk viscosities without significantly altering the overall expansion history. In particular, \citet{2012PhRvD..86h3501V} found that viscous dark matter is permitted to have a bulk viscosity up to $\zeta_0 \sim 10^7$~\si{Pa.s} at the background level. However, including bulk viscosity effects on the linear growth of dark matter perturbations leads to much tighter constraints. The effective sound speed, defined as $c_{\mathrm{eff}}^{2} = \delta\Pi/\delta\rho$, introduces additional scale-dependent damping terms in the perturbation equations. These terms lead to a strong suppression of small-scale structure formation, yielding constraints as stringent as $\zeta_0 \lesssim 10^{-3}~\si{Pa.s}$ when both the linear and nonlinear perturbations are considered \citep{2012PhRvD..86h3501V, 2014PhRvD..90l3526V}.

Notably, using PPS data alone yields $\zeta_0 = (0.72 ^{+1.35}_{-0.72}) \times 10^6$~\si{Pa.s} for the flat geometry, with the lower uncertainty bound approaching zero viscosity ($\zeta_0 \rightarrow 0$). We observe the same behavior when combining PPS with the DESI dataset, which gives $\zeta_0 = (0.47^{+0.89}_{-0.47})\times 10^6$~\si{Pa.s}. In contrast, non-flat scenarios produce significantly higher values: $\zeta_0 = (2.04^{+3.46}_{-1.99}) \times 10^6$~\si{Pa.s} (PPS) and $\zeta_0 = (7.46^{+3.60}_{-3.64}) \times 10^6$~\si{Pa.s} (Base~1). When combining both SH0ES calibration into PP (PPS) and the R22 prior, we obtain $\zeta_0 = (3.16^{+3.17}_{-2.43}) \times 10^6$~\si{Pa.s} in alignment with the results from the flat-case analysis. This pattern suggests that local measurements may indirectly constrain viscous effects in cosmic expansion.

\subsection{Free $m$ scenario}

\begin{figure}[h!]
    \centering
    \includegraphics[width=0.45\textwidth]{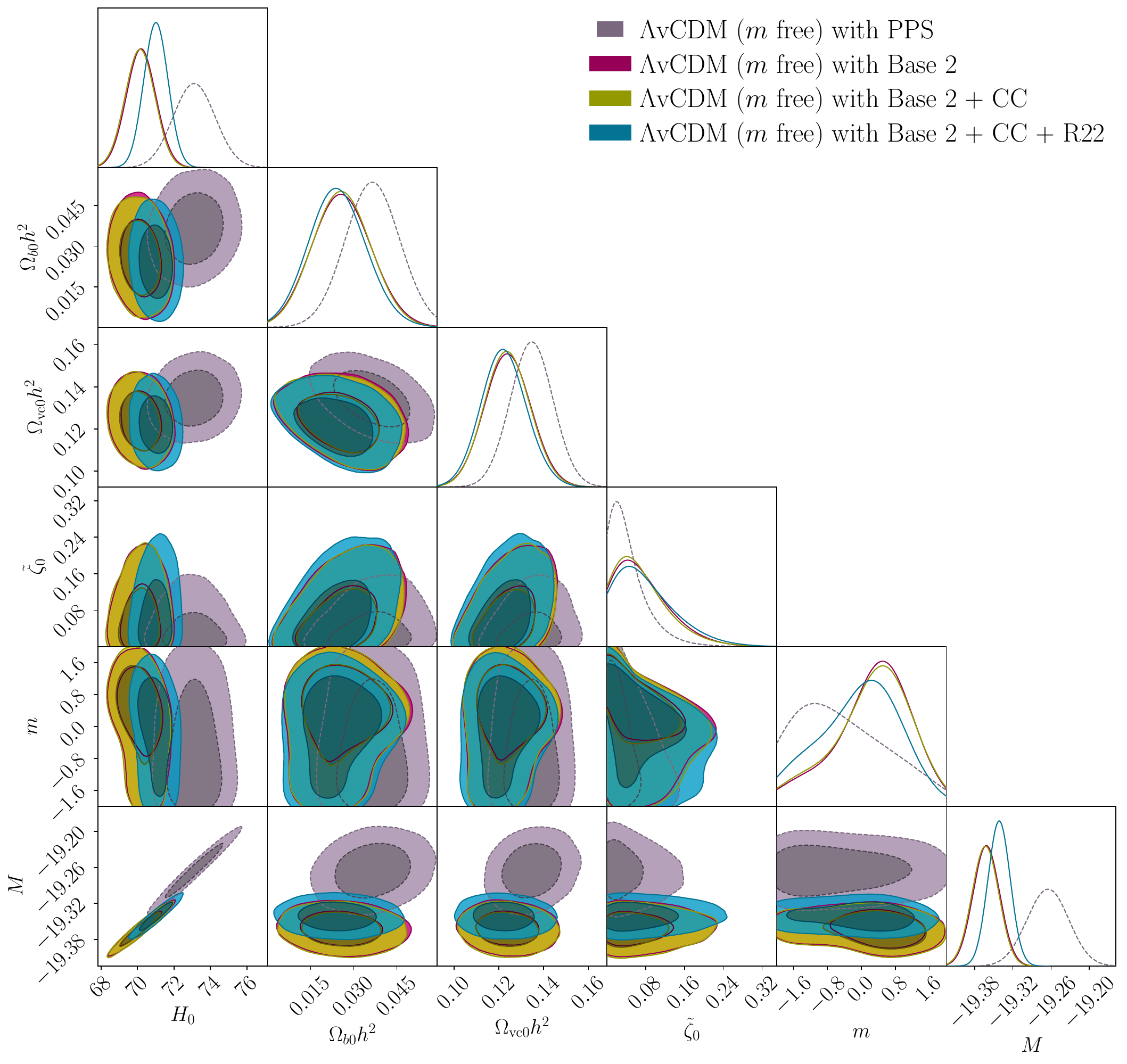}
    \includegraphics[width=0.45\textwidth]{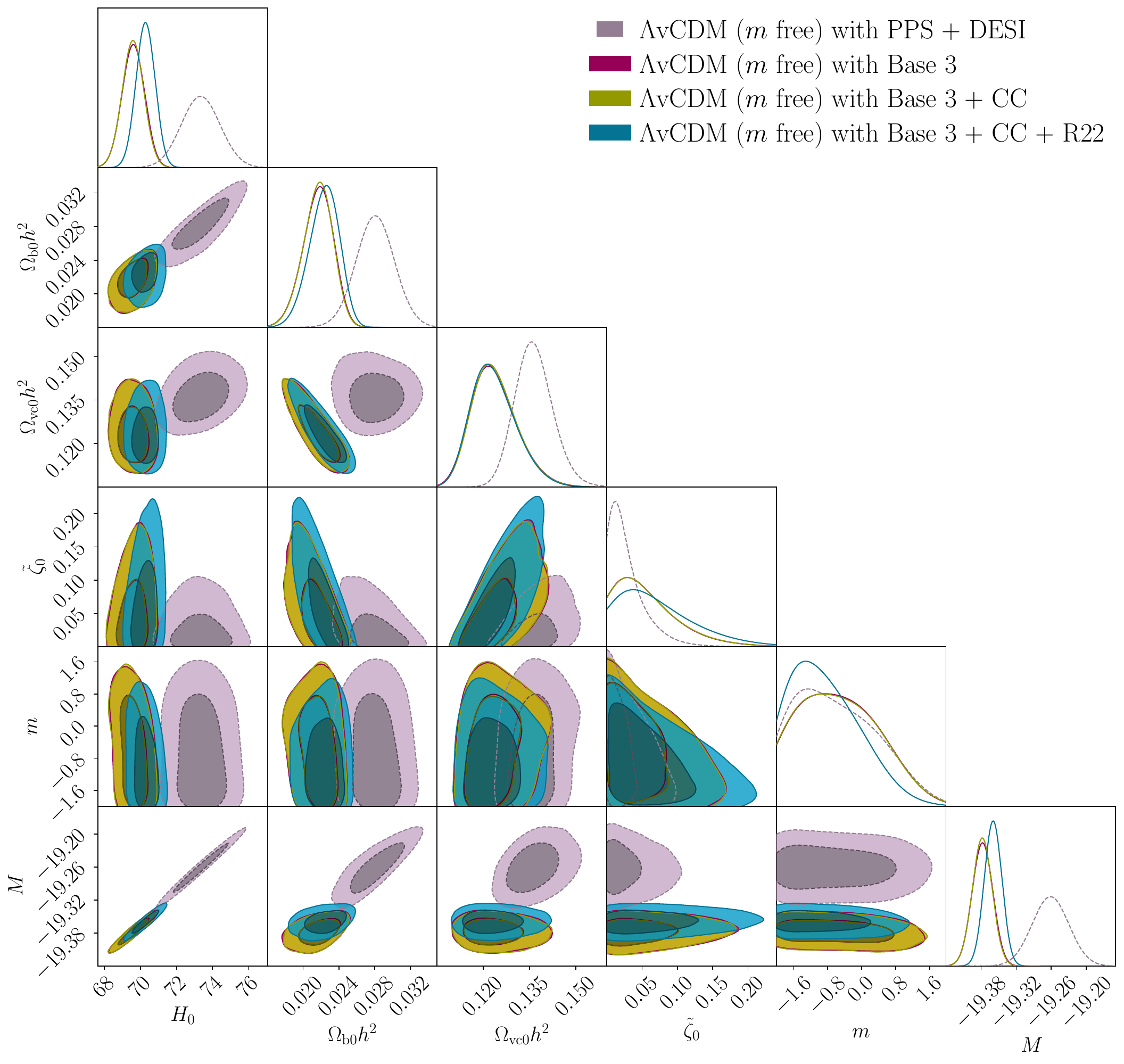}
    \caption{Cosmological parameter constraints for a flat viscous universe with free $m$. Top panel: PPS, Base~2, CC, and R22 datasets. Bottom panel: PPS + DESI, Base~3, CC, and R22 datasets. Contours show 2D confidence regions ($68\%$ and $95\%$ CL) and their 1D marginalized posterior distributions.}
    \label{fig:flat-LvCDM_mfree_JOIN_BASE1-2}
\end{figure}

\begin{figure}[h!]
\centering
   \includegraphics[width=0.45\textwidth]{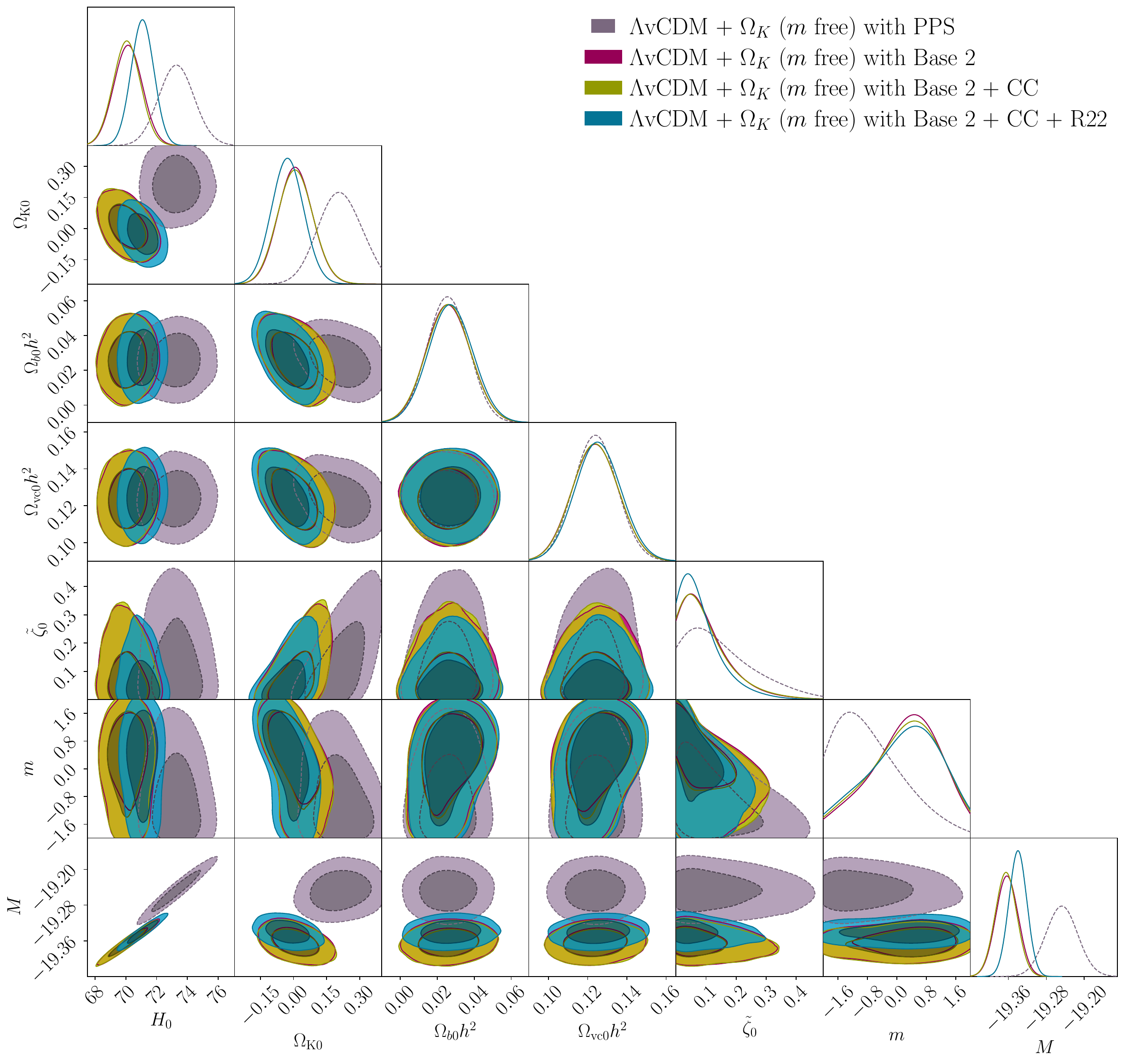}  
   \includegraphics[width=0.45\textwidth]{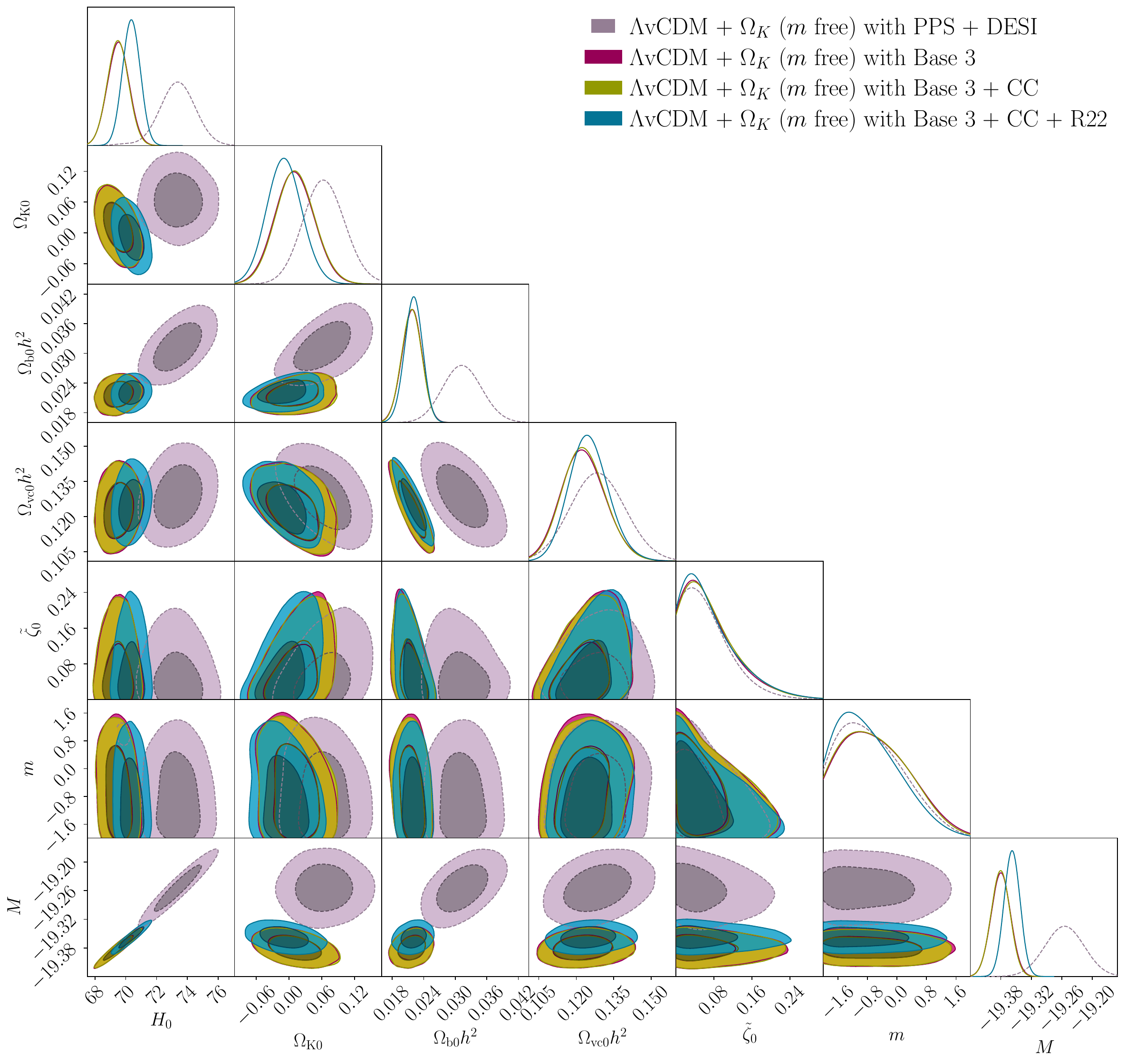}
\caption{Cosmological parameter constraints for a non-flat viscous universe with $m$ free. Top panel: PPS, Base~2, CC, and R22 datasets. Bottom panel: PPS + DESI, Base~3, CC, and R22 datasets. Contours show 2D confidence regions ($68\%$ and $95\%$ CL) and their 1D marginalized posterior distributions.}
\label{fig:LvCDM+OK_mfree_JOIN_BASE1-2}
\end{figure}

We extend our analysis to the more general case, where  $m$ is treated as a free parameter. Using the observational datasets Base~1, Base~2, and Base~3, combined with CC data and the R22 prior, we constrained both the flat and non-flat scenarios. The results are summarized in Table \ref{table:results_mfree} at the $68\%$ CL. Additionally, we present the parameter constraints in Fig.~\ref{fig:flat-LvCDM_mfree_JOIN_BASE1-2} (flat universe) and Fig.~\ref{fig:LvCDM+OK_mfree_JOIN_BASE1-2} (curved universe),  which display 2D confidence contours at $68\%$ and $95\%$ CL, as well as the corresponding 1D marginalized posterior distributions for the free parameters (see Table~\ref{table:free_parameters-priors}).

As in the $m=0$ case, Table \ref{table:results_mfree} shows that the datasets significantly affect the constraints on spatial curvature. The PPS sample alone does not place tight limits on the curvature, yielding $\Omega_{\rm{K0}} = 0.205^{+0.103}_{-0.094}$ ($1.99\sigma$). This is consistent within $1\sigma$ with \citep{2022ApJ...938..110B}, where $\Omega_{\rm{K0}} = 0.07^{+0.14}_{-0.13}$ \footnote{\url{https://github.com/PantheonPlusSH0ES/DataRelease}}. On the other hand, Base~1 alone shows a mild preference for an open universe, with ($\Omega_{\rm{K}0} =  0.21^{+0.13}_{-0.12}$), corresponding to $1.62\sigma$ deviation from flatness. However, including the R22 prior restores consistency with spatial flatness within $1 \sigma$, yielding

\begin{equation}
    \Omega_{\rm{K0}} = 0.026^{+0.087}_{-0.082} \quad (\mathrm{Base~1 + CC + R22}).
\end{equation}

\noindent
Further incorporating the PPS sample leads to significantly tighter constraints,

\begin{subequations}
\begin{align}
    \Omega_{\rm{K0}} &= 0.007^{+0.079}_{-0.077} \quad (\mathrm{Base~2}),\\
    \Omega_{\rm{K0}} &= 0.008^{+0.078}_{-0.081} \quad (\mathrm{Base~2 + CC}).
\end{align}
\end{subequations}

\noindent
This behavior persists with the inclusion of DESI data, yielding

\begin{subequations}
\begin{align}
    \Omega_{\rm{K0}} &= 0.008^{+0.037}_{-0.035} \quad (\mathrm{Base~3}),\\
    \Omega_{\rm{K0}} &= 0.010^{+0.036}_{-0.035} \quad (\mathrm{Base~3 + CC}),\\
   \Omega_{\rm{K0}} &= -0.009 \pm 0.032 \quad (\mathrm{Base~3 + CC + R22}). 
\end{align}
\end{subequations}

\noindent
Notably, when treating $m$ as a free parameter in combination with local measurements, we obtain the most stringent curvature constraints, further supporting spatial flatness. 

We find that the PPS yields estimates of $H_0$ consistent with SN calibrations using Cepheids \citep{2022ApJ...938..110B, 2022ApJ...934L...7R} across all scenarios considered. Specifically, for both $m=~0$ and free $m$ scenarios, using PPS alone, we obtain 

\begin{subequations}
    \begin{align}
        H_0 &= 73.1 \pm 1.1~\si{km.s^{-1}.Mpc^{-1}} \quad (\Lambda\mathrm{vCDM}), \\
        H_0 &= 73.3 \pm 1.2~\si{km.s^{-1}.Mpc^{-1}} \quad (\Lambda\mathrm{vCDM} + \Omega_{\mathrm{K}}).
    \end{align}
\end{subequations}

\noindent
Notably,  we see no significant shifts in $H_0$ values between $m=0$, free $m$, and $\tilde{\zeta}_0 = 0$ scenarios across all datasets (see Tables \ref{table:results_m=0}, \ref{table:results_mfree}, and \ref{table:model_LCDM-LCDM+OK}). 

Regarding the present-day bulk viscosity, computed from Eq. (\ref{eq:zeta0_tilde}), our results  yield a consistent value of  $\zeta_0 \sim 10^6~\si{Pa.s}$ for both flat and non-flat geometries (analogous to the $m=0$ case), in agreement with \citep{2017MPLA...3250026N, 2022Symm...14.1866C}.  In a flat universe, we find a shift toward higher values, ranging from $\zeta_0 = (0.51^{+1.10}_{-0.47}) \times 10^6$ (PPS + DESI)  to $\zeta_0 = (1.17^{+1.87}_{-1.13})\times 10^6$~\si{Pa.s} (Base~1 + CC), and up to  $\zeta_0 =  (1.97^{+2.88}_{-1.93})~\times 10^6$~\si{Pa.s} for Base~2 + CC + R22.  Including spatial curvature as a free parameter leads to even higher values, reaching $\zeta_0 = (8.03^{+8.80}_{-6.89}) \times 10^6$~\si{Pa.s} for Base~1, but decreasing to $\zeta_0 = (1.30^{+2.73}_{-1.22}) \times 10^6$~\si{Pa.s} for Base~3 + CC + R22. We find that all measured values of the bulk viscosity ($\zeta_0 >0$) are in full agreement with the second law of thermodynamics. This positivity requirement,  $\zeta_0 \geq 0$ is essential to ensure non-negative entropy production in an expanding universe \citep{1972gcpa.book.....W}.

From Figs.~\ref{fig:flat-LvCDM_mfree_JOIN_BASE1-2} and \ref{fig:LvCDM+OK_mfree_JOIN_BASE1-2}, we observe a moderate negative correlation between $m$ and $\tilde{\zeta}_0$ (e.g., $\rm{corr} = -0.43$ for PPS alone and $\rm{corr} =-0.52$ for Base~2 + CC) as well as an anti-correlation between $m$ and $\Omega_{\rm{K0}}$ ($\rm{corr} \sim -0.2$ for PPS alone and $\rm{corr} \sim -0.5$ for the other datasets). In contrast, $\tilde{\zeta}_0$ and $\Omega_{\rm{K0}}$ exhibit a positive correlation ($\rm{corr} \sim 0.5$) across all datasets in the free $m$ case. 

Finally, the sign of $m$ in Eqs.~\ref{eq:ansatz_zeta} -- \ref{eq:zeta0_tilde} determines the evolution of viscosity. Negative values ($m < 0$) indicate that viscosity increases as density decreases, meaning it grows with cosmic expansion. In contrast, positive values ($m>0$) correspond to viscosity that was stronger in the early Universe and weakens during expansion. The $m=0$ case represents constant viscosity. Our results (Table~\ref{table:results_mfree}) show that this behavior depends on the dataset. For flat geometry,  PPS alone strongly favors $m < 0$ (e.g., $m = -1.11^{+1.39}_{-0.82}$), while several combined datasets yield $m > 0$, although this trend may change when DESI data are included. A non-flat universe amplifies this effect, driving $m$ to more negative values (e.g., $m = -1.27^{+1.10}_{-0.72}$ for PPS alone) and yielding  $m< -0.4$ for Base~1 (+ CC), suggesting curvature-dependent viscous behavior. However, including PPS and the R22 prior shifts $m$ to positive values. On the other hand, when we include DESI data, the results consistently favor $m<0$ in both  flat and non-flat scenarios, implying that bulk viscosity grows with cosmic expansion.

\subsection{Model selection}

In the previous sections, we analyzed the observational constraints of various bulk viscous models by computing their respective $\chi^2_{\rm{min}}$ values. To assess their efficiency relative to the $\Lambda$CDM model, we evaluate the  AIC, BIC, and DIC information criteria as defined in Section \ref{sec:methodology-dataset} (Eqs. (\ref{eq:AIC}) -- (\ref{eq:DIC})). To complement these criteria, we employ Bayesian evidence, which incorporates prior knowledge for more nuanced model comparisons. In Table~\ref{table:summary_IC_BayesFactor}, we present the $\Delta$IC values relative to $\Lambda$CDM (see Eq. (\ref{eq:Delta_IC})), along with the Bayesian evidence and the Bayes factor for all viscous models.

\subsubsection{Information criteria}

From Table~\ref{table:summary_IC_BayesFactor}, we see that both PPS alone and its combinations with DESI data yield $\Delta\rm{AIC} < 0$ in the four scenarios. For the $m = 0$ case, the flat scenario yields $\Delta\mathrm{AIC} > 0$ for the Base~1 and Base~2 combinations, whereas PPS and the datasets including DESI provide $\Delta\mathrm{AIC} < 0$, indicating weak evidence for the $\Lambda$CDM model. Specifically, the viscous models show a positive preference with the Base~1(+ CC) datasets. However, this preference diminishes upon incorporating the PPS and R22 prior (e.g., $\Delta$AIC = 1.74 for Base~1 + CC + R22; $\Delta$AIC = 0.62 for Base~2 + CC + R22), but  Base~2 alone maintains positive evidence. The $\Lambda$vCDM + $\Omega_{\rm{K}}$ models show stronger evidence arguing against $\Lambda$CDM for the Base~1 dataset and positive evidence for the Base~1 + CC dataset. In contrast, the $\Lambda$CDM model becomes weakly preferred when either PPS or R22 prior is included (e.g., $\Delta\rm{AIC} = -0.90$ for Base~1 + CC + R22; $\Delta\rm{AIC} = -0.29$ for Base~2; $\Delta\rm{AIC} = -0.82$ for Base~3 and Base~3~+~CC). For the free $m$ case, the $\Lambda$vCDM models favor the viscous scenario only when using the Base~1 dataset and its combination with CC, with $\Delta\mathrm{AIC} = 6.53$ and $6.63$, respectively. However, this preference weakens when the R22 prior is included (e.g., $\Delta\mathrm{AIC} = 1.07$ for Base~1~+~CC~+~R22) and vanishes when the PPS sample or DESI data are incorporated, 
for which $\Delta\rm{AIC} < 0$, indicating a preference for $\Lambda$CDM. In contrast, the $\Lambda$vCDM$~+~\Omega_{\rm{K}}$ models yield $\Delta\rm{AIC} > 2$ for the Base~1 and Base~2 combinations, whereas including DESI again leads to $\Delta\mathrm{AIC} < 0$.

Regarding the BIC, all viscous scenarios yield $\Delta\rm{BIC} < 0$, favoring the $\Lambda$CDM model (see Table~\ref{table:summary_IC_BayesFactor}). The $\Delta$BIC  values impose a strong penalty on viscous models in nearly all cases. For $m=0$,  the $\Lambda$vCDM + $\Omega_{\rm{K}}$ model yields $\Delta\rm{BIC} < -11$ when including the R22 prior and PPS, indicating very strong evidence against this curved viscous model.  Even without these local measurements, we still find positive evidence against the $\Lambda$vCDM + $\Omega_{\rm{K}}$ models (e.g., $\Delta$BIC $= -4.78$ with Base~1). For flat scenarios, the Base~2 + CC + R22 constraints provide positive evidence against $\Lambda$vCDM ($\Delta\rm{BIC} = -4.83$), whereas analyses of these priors yield only weak evidence ($\Delta$BIC $=-0.84$ for Base~1 and $\Delta\rm{BIC} =-0.65$ for Base~1 + CC). For the free $m$ case, the $\Lambda$vCDM + $\Omega_{\rm{K}}$ model shows very strong evidence against this curved viscous scenario across all datasets ($\Delta\rm{BIC} < - 11$); in particular, $\Delta\rm{BIC} = -21.97$ for PPS alone. However, this evidence weakens when omitting the local measurements (R22 prior and PPS sample), yielding only positive evidence against the viscous models ($\Delta\rm{BIC} = -4.25$ for Base~1 and $\Delta\rm{BIC} = -4.19$ for Base~1 + CC).

The DIC analysis reveals distinct patterns across model configurations. The four scenarios exhibit mixed support depending on the dataset. The flat model with constant viscosity shows values ranging from $\Delta\rm{AIC} = -0.6$ (PPS) to $\Delta$AIC = 3.31 (Base~1+CC), indicating weak to positive evidence against $\Lambda$CDM.   In contrast,  the curved models show stronger evidence  against $\Lambda$CDM, particularly for the Base~1 and Base~1 + CC datasets, a pattern that persists when $m$ is free. The inclusion of PPS and R22 prior systematically reduces support for viscous models across all scenarios. Notably, the $\Lambda$vCDM + $\Omega_{\rm{K}}$ model with free $m$ shows the strongest evidence  against $\Lambda$CDM, reaching $\Delta$DIC = 15.53 for Base~2.

\subsubsection{Bayes factor} 

To complement the information criteria, we also employed Bayesian evidence, which provides more nuanced model comparisons by incorporating prior knowledge. While AIC and BIC apply fixed penalties based on parameter counts, DIC refines this by using the effective number of parameters constrained by the data via posterior distributions, Bayesian evidence inherently penalizes parameters that are unconstrained by data through marginalization over the prior space.

We estimated the natural logarithm of the Bayesian evidence for the bulk viscous models relative to $\Lambda$CDM using the \texttt{MCEvidence} code \citep{2017arXiv170403472H, 2017PhRvL.119j1301H}\footnote{The code is publicly available on \url{https://github.com/yabebalFantaye/MCEvidence}}. By reusing existing MCMC chains from parameter inference, we computed Bayes factors for various model-dataset combinations. We interpret the results using the Jeffreys scale \citep{jeffreys1961theory, Kass01061995}. Different conventions exist for classifying the strength of evidence in the Jeffreys scale, but here we follow the one proposed by \citep{2017PhRvL.119j1301H}, as summarized in Table \ref{table:strength_evidence}. 

Following the procedures  in the literature \citep{2017arXiv170403472H, 2017PhRvL.119j1301H}, we computed
$\ln \mathcal{B}_{0i}$, where $0$ refers to the benchmark $\Lambda$CDM model and $i$ indexes each of the four bulk viscous models. The results for all model-dataset combinations are summarized in Table \ref{table:summary_IC_BayesFactor} and calculated as

\begin{equation}
    \ln \mathcal{B}_{0i} = ~\ln \mathcal{E}_{\Lambda\rm{CDM}} - \ln \mathcal{E}_{i}.
\end{equation}

From Table \ref{table:summary_IC_BayesFactor}, we find moderate evidence against both  $\Lambda$vCDM + $\Omega_{\rm{K}}$ scenarios (for both $m = 0$ and free $m$ cases), with $ 3 \leq \ln \mathcal{B}_{01} < 5 $ for most datasets. This evidence against models with curvature is particularly strong when we include DESI data, with $\ln \mathcal{B}_{01} > 5$ across all datasets. The most disfavored case occurs for $m=0$, under the Base~3 + CC + R22 dataset ($\ln\mathcal{B}_{0i} = 7.18$). Conversely, the $\Lambda$vCDM  models with $m=0$ show only weak evidence across all datasets (e.g., $\ln \mathcal{B}_{0i} = 1.34$ and $\ln \mathcal{B}_{0i} = 1.40$ for Base~2 + CC + R22 and Base~2, respectively), while models with free $m$ are inconclusive.

\subsection{Evolution of cosmological parameters}

\begin{figure}[h!]
    \centering
    \includegraphics[width=0.85\linewidth]{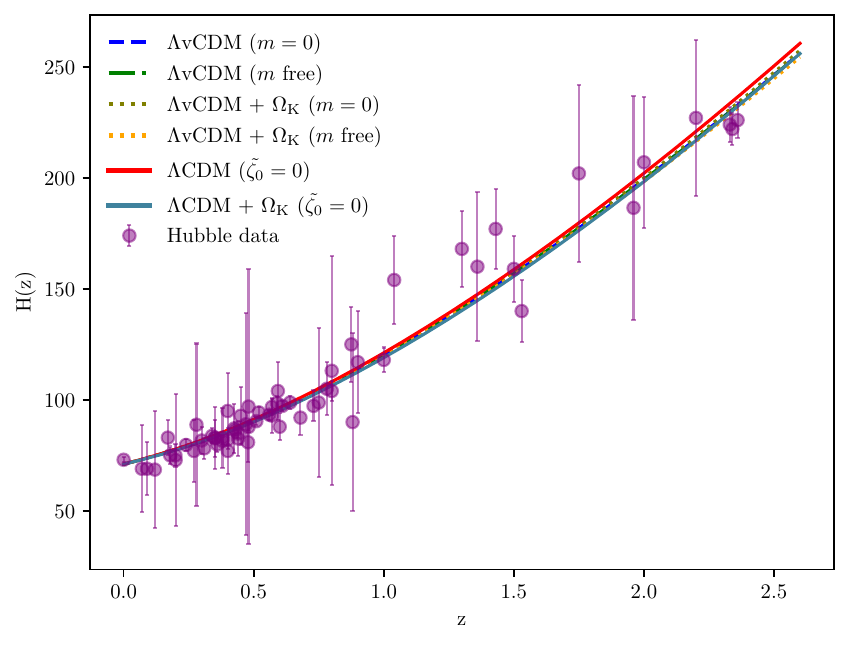}
    \caption{Evolution of Hubble function H(z) with redshift $z$. Comparison between OHD (purple points with error bars; $0 < z < 2.36$) and theoretical predictions. The solid red and green lines corresponds to the $\Lambda$CDM and $\Lambda$CDM + $\Omega_{\rm{K}}$ models, respectively. The dashed, dash-dotted, and dotted lines correspond to the $\Lambda$vCDM and $\Lambda$vCDM + $\Omega_{\rm{K}}$ models. The parameters constrained by the Base 2 + CC + R22 dataset.}
    \label{fig:plots_OHD-evol_H(z)_models}
\end{figure}

Figure~\ref{fig:plots_OHD-evol_H(z)_models}  shows the evolution of the Hubble parameter as a function of the redshift given in Eq.~(\ref{eq:evol_Hubble}). The model exhibits good agreement with observational Hubble data (OHD; combined CC + BAO + R22 data) at low redshifts ($z < 1.5$). However, it shows increasing tension with observations at higher redshifts.

\begin{figure}[h!]
    \centering
    \includegraphics[width=0.85\linewidth]{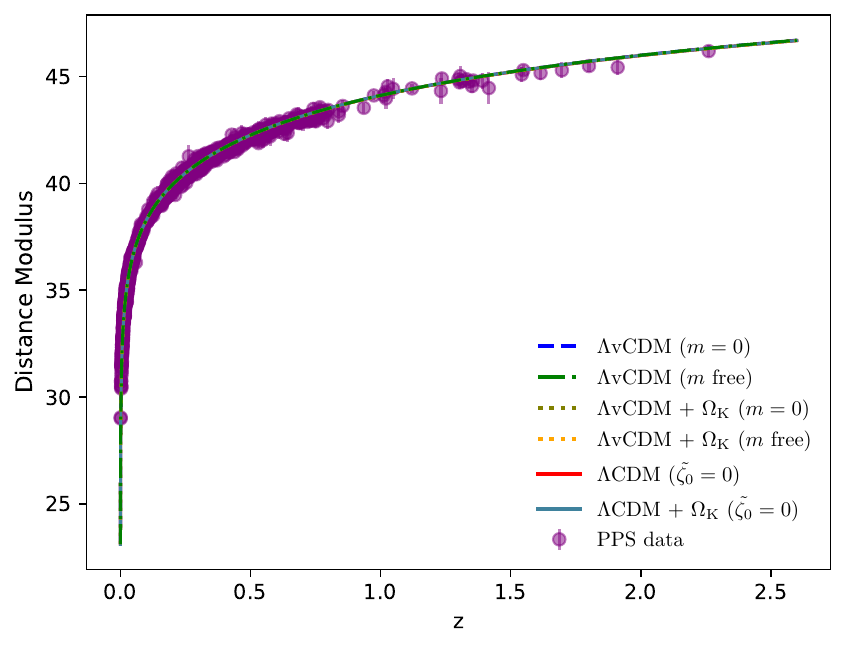}
    \caption{Comparison between the observed distance modulus from PPS compilation (purple points with error bars) within the redshift range $0.001 < z < 2.26$  and theoretical predictions. The solid red and green line corresponds to the $\Lambda$CDM and $\Lambda$CDM + $\Omega_{\rm{K}}$ models, respectively. While the dashed, dash-dotted, and dotted lines represent different cases of the $\Lambda$vCDM and $\Lambda$vCDM + $\Omega_{\rm{K}}$ models. The parameters constrained by the Base 2 + CC + R22 dataset.}
    \label{fig:plots_data-Dist_modulo_models}
\end{figure}

The distance modulus for SNe Ia derived from our model is given in Eq.~(\ref{eq:dist_modulus_model}). We compared these theoretical predictions with the PPS sample,  comprising 1701 light curves from 1550 distinct SNe Ia covering a broad redshift range (see Fig. \ref{fig:plots_data-Dist_modulo_models}). The model shows good agreement with the observational data.

\section{Conclusions}
\label{sec:conclusions}

In this work, we carried out an observational analysis of four bulk viscous cosmological models with a cosmological constant, where the bulk viscosity coefficient evolves according to a power law of $\zeta = \zeta_0 (\Omega_{\rm{vc}}/\Omega_{\rm{vc0}})^m$. We explored both constant viscosity ($m=0$) and variable viscosity ($m$ free) scenarios in flat and curved geometries. Using a Bayesian MCMC approach, we constrained these models with Hubble parameter measurements, $H(z)$, from CC and an independent compilation of BAO measurements, BAO data from DESI DR2, SNe Ia data (PP/PPS), and the R22 prior. Our main conclusions  are as follows.

The Hubble tension persists; however, it is partially alleviated across all investigated scenarios when local measurements are included. For the flat universe cases, we found $H_0 = 67.94^{+0.97}_{-0.92}$~\si{km.s^{-1}.Mpc^{-1}} (Base~1, $m=0$),  $H_0 = 70.39^{+0.75}_{-0.73}$~\si{km.s^{-1}.Mpc^{-1}} (Base~2, $m=0$), and $H_0 = 69.53 \pm 0.56$~\si{km.s^{-1}.Mpc^{-1}} (Base~3, $m=0$). When combining datasets with the R22 prior, the tension reduces to approximately  $1\sigma$, yielding  $H_0 = 69.92^{+0.71}_{-0.74}$~\si{km.s^{-1}.Mpc^{-1}} (Base~1 + CC + R22), $H_0 = 71.05^{+0.62}_{-0.60}$~\si{km.s^{-1}.Mpc^{-1}} (Base~2 + CC + R22), and $H_0 = 70.17^{+0.49}_{-0.51}$~\si{km.s^{-1}.Mpc^{-1}} (Base~3 + CC + R22). This reduction in tension persists across all tested configurations: spatial geometry (flat/curved), power-law bulk viscosity evolution ($m=0$/free $m$), and null present-day bulk viscosity ($\tilde{\zeta}_0 = 0$) scenarios.

For the non-flat scenarios, the models initially favor negative curvature (corresponding to $\Omega_{\rm{K0}} > 0$) at more than $2\sigma$ significance in some cases, suggesting a moderate preference for an open universe. However, when the PPS sample and the R22 prior are included, the constraints shift toward spatial flatness ($\Omega_{\rm{K0}} \approx 0$), consistent within $1\sigma$. This consistency with a flat universe is also achieved, with even tighter uncertainties, when DESI data is incorporated into any of the dataset combinations.

Our results indicate that the present-day bulk viscosity maintains a consistent value of $\zeta_0 \sim 10^6$~\si{Pa.s} across all investigated scenarios.  This aligns with the known hierarchy of cosmological constraints, where background expansion permits viscosities several orders of magnitude larger than those allowed by the suppression of small-scale structure formation, as discussed in previous studies \citep{2012PhRvD..86h3501V,2014PhRvD..90l3526V}. Furthermore, the positive values of $\zeta_0$ preferred by our data are physically consistent, satisfying both the second law of thermodynamics within the Eckart framework and remaining compatible with cosmological observations. 
Specifically, values range from $\zeta_0 = 0.47^{+0.89}_{-0.47} \times 10^6$~\si{Pa.s} (PPS + DESI, flat case) to $\zeta_0 = 7.46^{+3.60}_{-3.64} \times 10^6$~\si{Pa.s} (Base~1, non-flat). On the other hand, the viscosity evolution parameter $m$ remains weakly constrained. Using PPS alone favors negative values ($m < 0$), corresponding to viscosity that increases as density decreases (i.e., growing with cosmic expansion). This preference is also favored by the inclusion of DESI data, which consistently favors $m<0$ across both flat and non-flat scenarios, suggesting that bulk viscosity becomes more significant at later times. In contrast, joint analyses without DESI tend to prefer $m>0$, indicating that viscosity was more dominant in the early Universe.

Model selection criteria present a nuanced picture. We found that the AIC and DIC criteria favor viscous models in some observational datasets (e.g., $\Delta$AIC > 4 and $\Delta$DIC $> 6$ for Base~1(+CC)), while the BIC strongly prefers the benchmark $\Lambda$CDM model, as it imposes a more severe penalty for additional parameters. We observed that Bayesian evidence calculations further show moderate evidence against non-flat viscous scenarios ($3 \leq \ln \mathcal{B}_{0i} < 5$)  for most datasets. This evidence against models with curvature is particularly strong when we include DESI data, with $\ln \mathcal{B}_{0i} > 5$ across all datasets. These results suggest that although bulk viscosity can moderately improve cosmological fits, it neither resolves the $H_0$ tension nor outperforms $\Lambda$CDM when local measurements are included.

Our analysis reveals large uncertainties in $m$ and $\zeta_0$ when using $H(z)$ data, SNe Ia samples, and the R22 prior, demonstrating that these datasets cannot definitively distinguish between viscous scenarios. Consequently, the standard $\Lambda$CDM model remains the most parsimonious description. 
Nevertheless, viscous cosmologies still retain significant potential as physically motivated alternative models meriting investigation. To achieve more conclusive results, we aim to extend our investigation to incorporate CMB observations.

\begin{acknowledgements}
This research is part of the Ph.D. thesis of R.N. Villalobos  and is supported by the Agencia Nacional de Investigación y Desarrollo (ANID) through the Chilean National Doctoral Fellowship Nº 21212268. Additionally, this investigation was made possible thanks to funding from the Beca de Arancel y Manutención of the Dirección de Postgrados y Postítulos at the Universidad de La Serena (ULS), along with financial contributions from the Astronomy Doctoral Program at ULS. This work is partially supported by ANID Chile through FONDECYT Grant Nº 1220871 (R.N.V. and Y.V) and FONDECYT Grant No. 1250969 (N.C.). The authors wish to thank the FIULS 2030 project 18ENI2-104235 -- CORFO for providing computing resources. We thank the referee for their useful comments and suggestions. Lastly, R.N.V. dedicates this work to their children and Leonardo, whose unwavering love and support have been their  strength throughout this academic journey.
\end{acknowledgements}

\bibliographystyle{aa} 
\bibliography{references}

\begin{appendix}

\FloatBarrier
\onecolumn

\section{Observational data}
\label{Appendix:data}

\subsection{H(z) data}
\label{Appendix:OHD_data}

\begin{table}[h!]
\caption{33 H(z) measurements used in this  analysis, in units of \si{km.s^{-1}.Mpc^{-1}},  obtained with the differential age method and their associated uncertainties.}
\label{table:CC_data}
\centering
\begin{tabular}{llcllcllc}
\midrule\midrule
    z & H(z) $\pm \,\sigma_H$ & References &  z & H(z) $\pm \,\sigma_H$ & References  &   z & H(z) $\pm \,\sigma_H$ & References \\  
    \midrule
    0.07   & 69.0 $\pm$ 19.6 & 1 & 0.4    & 95 $\pm$ 17      & 3 & 0.80   & $113.1 \pm~25.8$ & 9  \\ 
    0.09   & 71 $\pm$ 12     & 2 & 0.4004 & 77.0 $\pm$ 10.2  & 5 & 0.8754 & 125 $\pm$ 17     & 4  \\ 
    0.12   & 68.6 $\pm$ 26.2 & 1 & 0.4247 & 87.1 $\pm$ 11.2  & 5 & 0.88   & 90 $\pm$ 40      & 3  \\ 
    0.17   & 83 $\pm$ 8      & 3 & 0.4497 & 92.8 $\pm$ 12.9  & 5 & 0.9    & 117 $\pm$ 23     & 7  \\ 
    0.1791 & 75 $\pm$ 4      & 3 & 0.47   & 89 $\pm$ 50      & 6 & 1.037  & 154 $\pm$ 20     & 4  \\ 
    0.1993 & 75 $\pm$ 5      & 4 & 0.4783 & 80.9 $\pm$ 9.0   & 5 & 1.3    & 168 $\pm$ 17     & 3  \\ 
    0.20   & 72.9 $\pm$ 29.6 & 1 & 0.48   & 97 $\pm$ 62      & 7 & 1.363  & $160.0 \pm~33.6$ & 10  \\ 
    0.27   & 77 $\pm$ 14     & 3 & 0.5929 & 104 $\pm$ 13     & 4 & 1.43   & 177 $\pm$ 18     & 3  \\ 
    0.28   & 88.8 $\pm$ 36.6 & 1 & 0.6797 & 92 $\pm$ 8       & 4 & 1.53   & 140 $\pm$ 14     & 3  \\
    0.3519 & 83 $\pm$ 14     & 4 & 0.75   & 98.8 $\pm$ 33.6  & 8 & 1.75   & 202 $\pm$ 40     & 3  \\
    0.3802 & 83.0 $\pm$ 13.5 & 5 & 0.7812 & 105 $\pm$ 12     & 4 & 1.965  & $186.5 \pm ~50.4$ & 10  \\
\midrule
\end{tabular}
\tablebib{(1)~\citet{2014RAA....14.1221Z}; (2) \citet{2003ApJ...593..622J}; (3)\citet{2005PhRvD..71l3001S}; (4) \citep{2012JCAP...08..006M}; (5) \citet{2016JCAP...05..014M}; (6) \citet{2017MNRAS.467.3239R}; (7) \citet{2010JCAP...02..008S}; (8) \citet{2022ApJ...928L...4B}; (9) \citet{2023ApJS..265...48J}; (10) \citep{2015MNRAS.450L..16M}.}
\end{table}

\begin{table}[!h]
\caption{30 H(z) measurements in units \si{km.s^{-1}.Mpc^{-1}}  derived from BAO analyses, including measurements from the galaxy distribution, the BAO signal in the $\rm{Ly}\alpha$ forest alone, and the Ly$\alpha$-Quasar cross-correlation.}
\label{table:HBAO_data}
\centering
\begin{tabular}{llcllcllc}
\midrule\midrule
    z & $\rm{H(z)} \pm ~\sigma_H$ & References &  z & $\rm{H(z)} \pm ~\sigma_H$ &  References &  z & $\rm{H(z)} \pm ~\sigma_H$ &  References \\ \midrule
    0.24 & $79.69 \pm ~2.65$ & 1  & 0.44 & $84.81 \pm ~1.83$ & 3  &  0.64 & $98.82 \pm ~2.98$ & 3 \\
    0.3  & $81.7 \pm ~ 6.22$ & 2  & 0.48 & $87.79 \pm ~2.03$ & 3  &  0.73 & $97.3 \pm ~7.0$   & 6 \\  
    0.31 & $78.18 \pm ~4.74$ & 3  & 0.51 & $90.4 \pm ~1.9$   & 5  &  0.8  & $104.0 \pm ~4.4$  & 8 \\ 
    0.34 & $83.80 \pm ~3.36$ & 1  & 0.52 & $94.36 \pm ~2.64$ & 3  &  1.0  & $118.0 \pm ~5.6$  & 8 \\ 
    0.35 & $82.7 \pm ~8.4$   & 4  & 0.56 & $93.34 \pm ~2.30$ & 3  &  1.5  & $159 \pm ~15$     & 8\\ 
    0.36 & $79.94 \pm ~3.38$ & 3  & 0.57 & $96.8 \pm ~3.4$   & 7a &  2.0  &  $207 \pm ~29$    & 8\\
    0.38 & $81.5 \pm ~ 1.9$  & 5  & 0.57 & $92.9 \pm ~7.8$   & 7b &  2.2  &  $227 \pm ~35$    & 8 \\
    0.40 & $82.04 \pm ~ 2.034$ & 3 & 0.59 & $98.48 \pm ~3.18$ & 3  &  2.33 & $224 \pm ~8 $     & 9 \\
    0.43 & $86.45 \pm ~3.68$ & 1  & 0.6  & $87.9 \pm ~6.1$   & 6  &  2.34 & $222 \pm ~7 $     & 10\\
    0.44 & $82.6 \pm ~7.8$   & 6  & 0.61 & $97.3 \pm ~2.1$   & 5  &  2.36 & $226 \pm ~8 $     & 11\\
\midrule
\end{tabular}
\tablebib{(1)\citet{2009MNRAS.399.1663G}; (2) \citet{2014MNRAS.439.2515O}; (3) \citet{2017MNRAS.469.3762W}; (4) \citet{2013MNRAS.435..255C}; (5) \citet{2017MNRAS.470.2617A}; (6) \citet{2012MNRAS.425..405B}; (7a) \citet{2014MNRAS.441...24A}; (7b) \citet{2014MNRAS.439...83A};  (8) \citet{2018MNRAS.477.1528W}; (9) \citet{2017AA...603A..12B}; (10) \citet{2015AA...574A..59D}; (11) \citet{2014JCAP...05..027F}.}
\end{table}

\subsection{DESI data}
\label{Appendix:DESI_data}

\begin{table}[h!]
  \centering
  \caption{BAO measurements from DESI DR2 used in this analysis, adapted from Table IV of \citet{2025PhRvD.112h3515A}.}
  \label{tab:desi_data}
  \begin{tabular}{lcccc}
    \midrule \midrule
    Tracer & $z_{\rm{eff}}$ & $D_{\rm{V}}(z)/r_{\rm{d}}$ & $D_{\rm{M}}/r_{\rm{d}}$ & $D_{\rm{H}}/r_{\rm{d}}$ \\
    \midrule
    BGS & 0.295 & $7.942 \pm 0.075$ & -- & --  \\
    LRG1 & 0.510 & -- & $13.588 \pm 0.167$ & $21.863 \pm 0.425$  \\
    LRG2 & 0.706 & -- & $17.351 \pm 0.177$ & $19.455 \pm 0.330$ \\
    LRG3+ELG1 & 0.934 & -- & $21.576 \pm 0.152$ & $17.641 \pm 0.193$  \\
    ELG2 & 1.321 & -- & $27.601 \pm 0.318$ & $14.176 \pm 0.221$  \\
    QSO & 1.484 & -- & $30.512 \pm 0.760$ & $12.817 \pm 0.516$ \\
    Ly$\alpha$ & 2.330 & -- & $38.988 \pm 0.531$ & $8.632 \pm 0.101$  \\
    \midrule
  \end{tabular}
\tablefoot{The tracers include: Bright Galaxy Sample (BGS), Luminous Red Galaxies (LRGs), Emission Line Galaxies (ELGs), Quasars (QSO), and the Lyman-$\alpha$ forest (Lya).}
\end{table}

\FloatBarrier
\twocolumn

\section{Case for $\tilde{\zeta}_0 = 0$: $\Lambda$CDM model}
\label{Appendix:LCDM_model}

We conduct a comparative MCMC analysis between the bulk viscous models, $\Lambda$vCDM and $\Lambda$vCDM + $\Omega_{\rm{K}}$, considering both constant ($m=0$) and variable viscosity ($m$ free). These models are evaluated against the Benchmark $\Lambda$CDM model (with $\tilde{\zeta}_0 = 0$). For quantitative model selection, we compute AIC, BIC, DIC, and $\ln \mathcal{E}_{i}$ for all models. To ensure transparency, we present the reference $\Lambda$CDM results (see Table~\ref{table:model_LCDM-LCDM+OK}).

\begin{figure}[h!]
    \centering
    \includegraphics[width=0.78\linewidth]{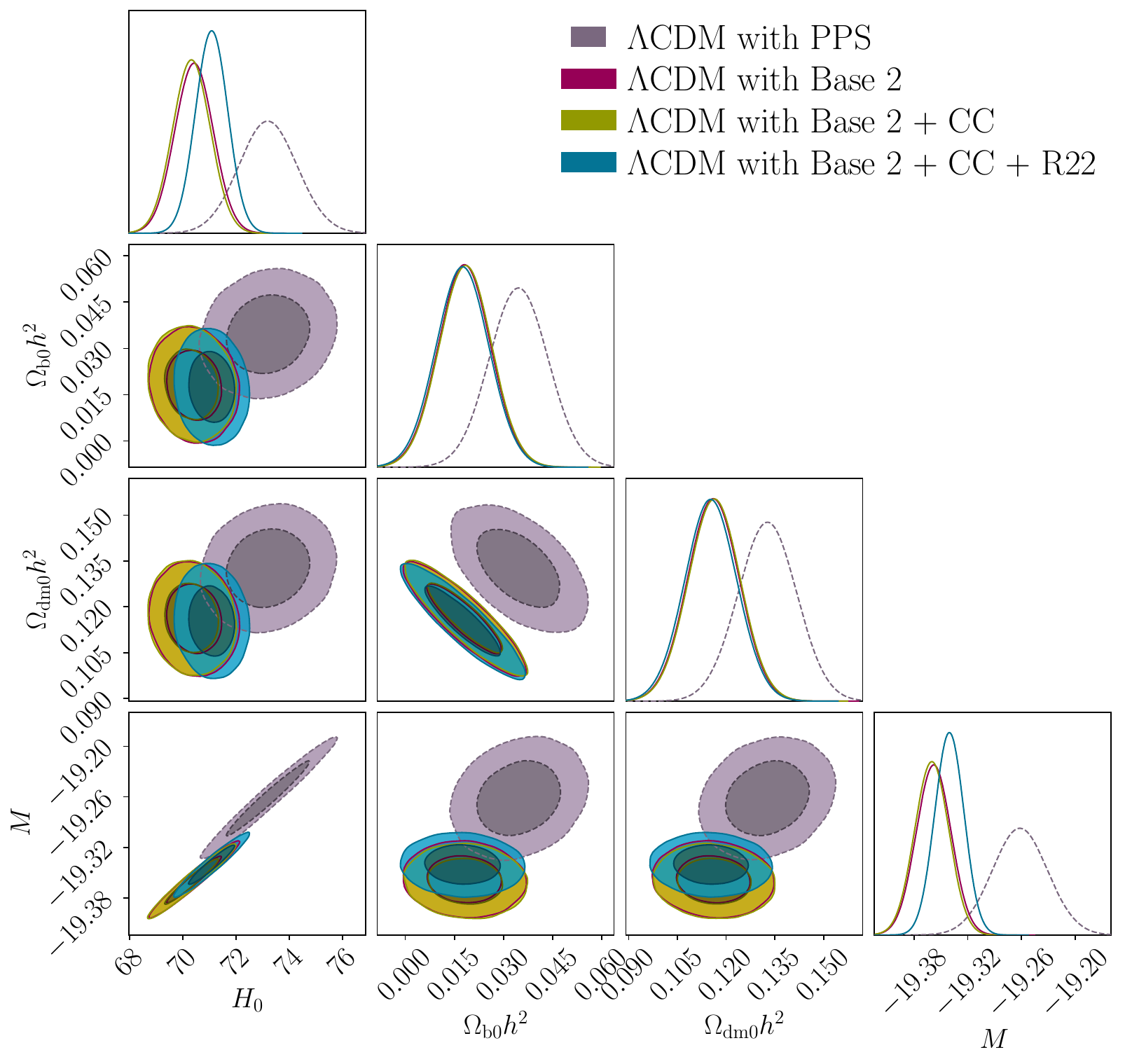}
    \includegraphics[width=0.78\linewidth]{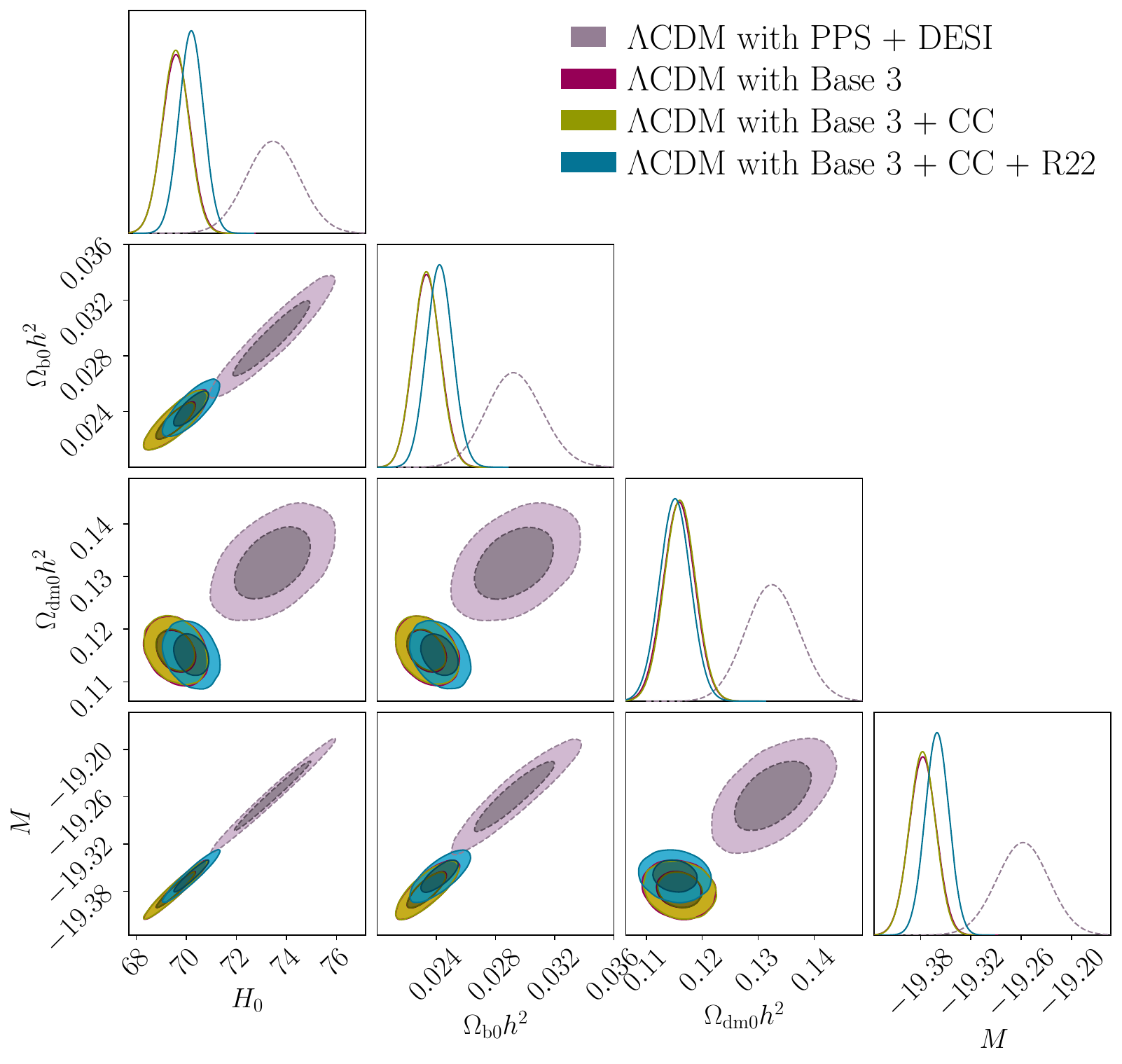}
    \caption{Cosmological parameter constraints for $\Lambda$CDM model with $\tilde{\zeta}_0 = 0$. Top panel: PPS, Base~2, CC, and R22 datasets. Bottom panel: PPS+DESI, Base~3, CC, and R22 datasets. Contours show 2D confidence regions ($68\%$ and $95\%$ CL) and their 1D marginalized posterior distributions.}
    \label{fig:plots_flatLCDM_models_Base1_Base2}
\end{figure}

\begin{figure}[h!]
    \centering
    \includegraphics[width=0.78\linewidth]{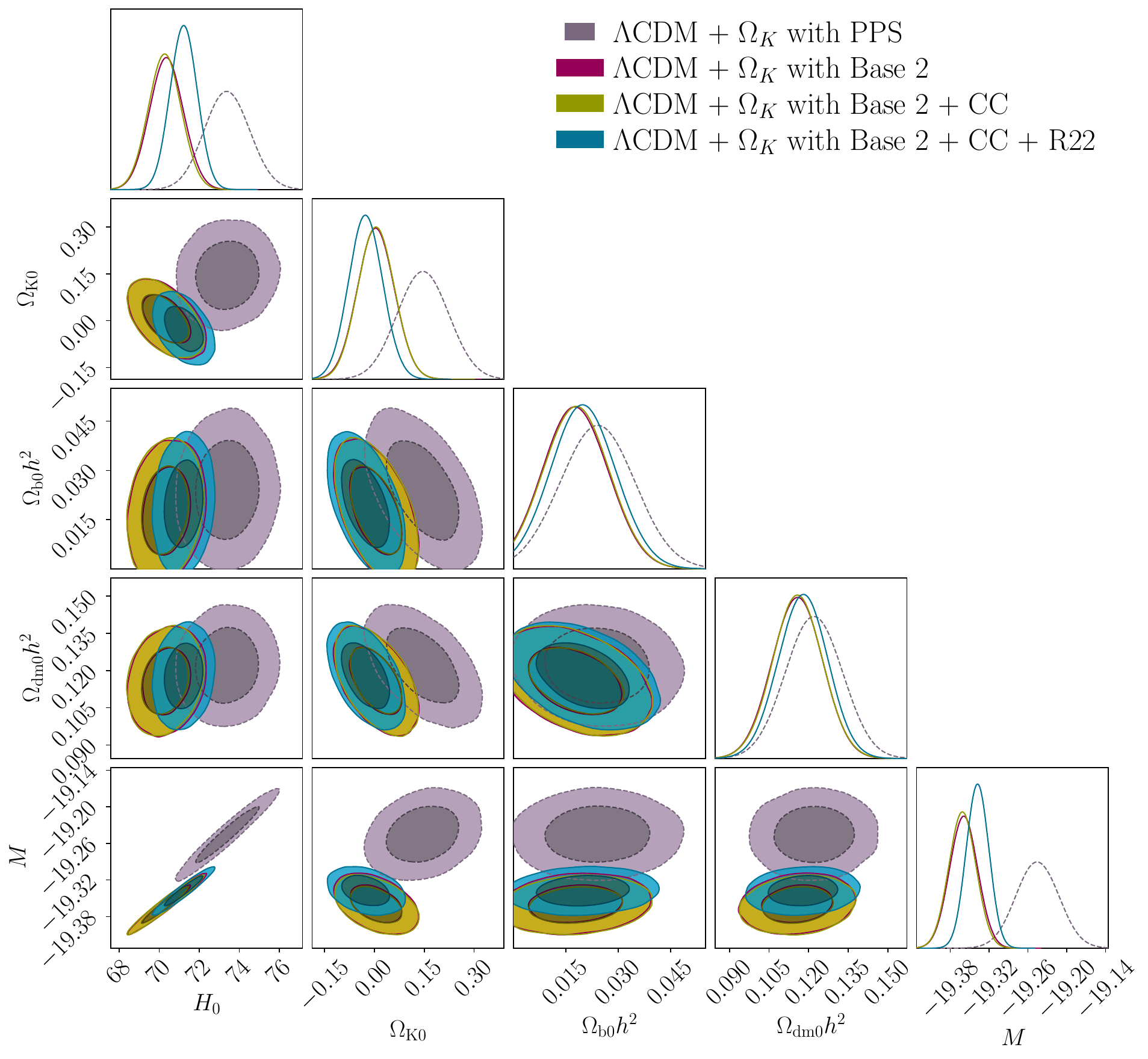}
    \includegraphics[width=0.78\linewidth]{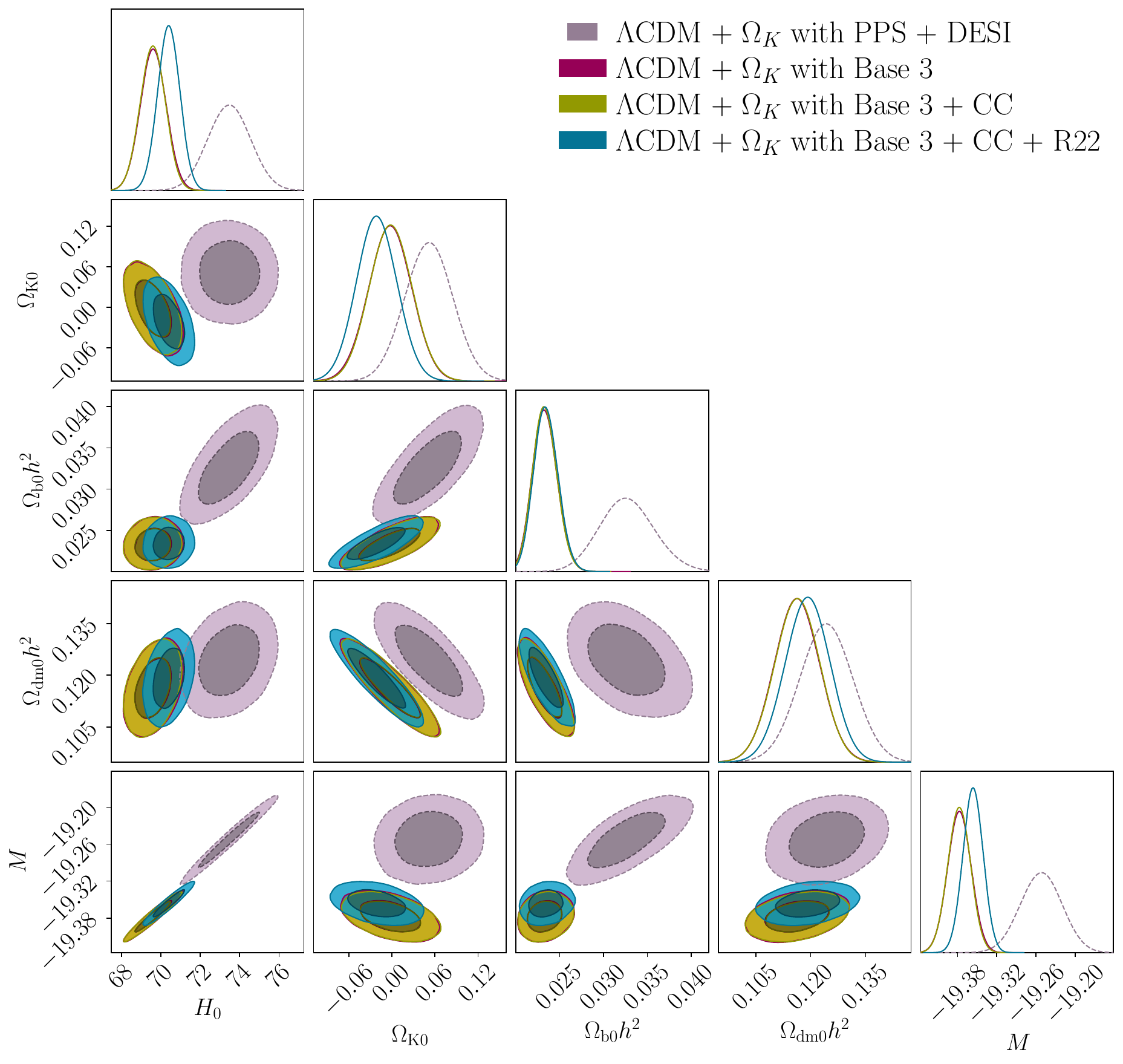}
    \caption{Cosmological parameter constraints for $\Lambda$CDM + $\Omega_{\rm{K}}$ model with $\tilde{\zeta}_0 = 0$. Top panel: PPS, Base~2, CC, and R22 datasets. Bottom panel: PPS+DESI, Base~3, CC, and R22 datasets. Contours show 2D confidence regions ($68\%$ and $95\%$ CL) and their 1D marginalized posterior distributions.}
\label{fig:plots_LCDM+OK_models_Base1_Base2}
\end{figure}

Figure~\ref{fig:plots_flatLCDM_models_Base1_Base2} shows the posterior distributions for the PPS dataset (purple-gray), while Figure~\ref{fig:plots_LCDM+OK_models_Base1_Base2} displays the results for the PPS+DESI dataset (purple). Both figures include the confidence regions from our MCMC runs for the other dataset combinations: Base~2, Base~2 + CC, and Base~2 + CC + R22 (dark magenta, green, and sea blue, respectively), alongside their Base~3 counterparts (same color scheme). In all cases, the chains converge rapidly, yielding nearly Gaussian uncertainties.

Table~\ref{table:model_LCDM-LCDM+OK} summarizes the best-fit parameters along with their $68\%$ CL uncertainties. We adopt the priors listed in Table~\ref{table:free_parameters-priors} for the $\tilde{\zeta}_0 = 0$ case, applying Gaussian priors for $\Omega_{\rm{dm}0}h^2$ and $\Omega_{\rm{b}0}h^2$, and  uniform priors for the remaining parameters ($H_0$ and $M$). We also report results for the curved $\Lambda\rm{CDM} + ~\Omega_{\rm{K}}$ extension in Table~\ref{table:model_LCDM-LCDM+OK}, where we assume a uniform prior for $\Omega_{\mathrm{K}0}$.

We quantify the relative evidence against the benchmark $\Lambda$CDM model by computing the following  $\Delta$ values:

\begin{align}
    &\Delta \rm{IC} = \rm{IC}_{\Lambda CDM} - \rm{IC}_{\Lambda CDM + \Omega_{\rm{K}}},\\
    &\ln \mathcal{B}_{ij} = \ln \mathcal{E}_{\Lambda \rm{CDM}} - \ln \mathcal{E}_{\Lambda\rm{CDM} + \Omega_{\rm{K}}}\,.
\end{align}

We find  that the Base~1 and Base~1 + CC datasets show strong evidence for spatial curvature in information criteria, with $\Delta\rm{AIC}>6$ and positive evidence in DIC (e.g., $\Delta\rm{DIC} = 5.83$ for  Base~1 + CC), though their BIC values demonstrate weaker support. Including R22 prior reverses this trend, resulting in negative $\Delta$IC values ($\Delta\rm{AIC} = -1.32$ for Base~1 + CC + R22; $\Delta\rm{BIC} = -6.73$; $\Delta\rm{DIC} = -0.64$). Across all datasets, Bayes factors consistently favor the flat $\Lambda$CDM model, reaching maximum evidence for the Base~3+CC+R22 dataset ($\ln\mathcal{B}_{0i}$ = 4.87).

\begin{sidewaystable*}
\caption{Summary of parameter constraints and model comparison between $\Lambda$CDM and curved $\Lambda$CDM at $68\%$ CL.}
\label{table:model_LCDM-LCDM+OK}
\centering
\begin{tabular}{lcccccccccccc}
\midrule \midrule
     \multirow{3}{*}{Datasets} & \multicolumn{10}{c}{$\Lambda$CDM ($\tilde{\zeta}_0 = 0$)}\\ \cmidrule{2-13}
   & $H_0$ & $\Omega_{\rm{K0}}$ &  $M$  & $\chi^2_{\rm{min}}$ & $\rm{AIC}_{\Lambda\rm{CDM}}$ & $\Delta$AIC & $\rm{BIC}_{\Lambda\rm{CDM}}$& $\Delta$BIC & $\rm{DIC}_{\Lambda\rm{CDM}}$ & $\Delta$DIC & $\ln \mathcal{E}_{0}$ & $\ln \mathcal{B}_{0i}$\\ 
   & (\si{km.s^{-1}.Mpc^{-1}}) & & &&\\
    \midrule
    PPS & $73.2\pm 1.1$ & -- &  $-19.262^{+0.032}_{-0.031}$ & 1452.03 & 1460.03  & 0 & 1481.68  & 0 & 1458.81 & 0 & $-4.49$ & 0 \\
    PPS + DESI & $73.4^{+1.1}_{-1.0}$ & -- & $-19.257^{+0.030}_{-0.032}$ & 1465.68 & 1473.68 & 0 & 1495.36 & 0 & 1473.60 & 0 & $-5.26$ & 0 \\
   \midrule
    Base 1 & $67.97^{+0.94}_{-0.91}$ &  -- & $-19.426\pm 0.026$ & 1427.36 & 1435.36 & 0 & 1456.92 & 0 & 1433.04 & 0 & $-4.62$ & 0 \\ 
    Base 1 + CC  & $68.05^{+0.89}_{-0.90}$& -- &  $-19.424\pm 0.025$ & 1442.53 & 1450.53 & 0 & 1472.17 &  0 & 1448.60 & 0 & $-4.72$ & 0 \\ 
        Base 1 + CC + R22  & $69.94^{+0.69}_{-0.72}$ & -- &   $-19.373^{+0.019}_{-0.020}$  & 1456.75  & 1464.75 & 0  & 1486.39 & 0 & 1462.85 & 0 & $-4.99$ & 0\\
    \midrule
    Base 2  & $70.41^{+0.72}_{-0.71}$ & -- &  $-19.358^{+0.020}_{-0.019}$ & 1498.52  & 1506.52 & 0 & 1528.24 & 0 & 1504.38 & 0 & $-4.94$ & 0\\
    Base 2 + CC  & $70.32^{+0.71}_{-0.70}$ &  -- &  $-19.360\pm 0.019$ & 1513.98 & 1521.98 & 0 & 1543.78 & 0 & 1519.83 & 0 & $-5.00$ & 0\\ 
        Base 2 + CC + R22* & $71.08^{+0.60}_{-0.61}$ & -- &   $-19.340\pm 0.016$ & 1518.76 &  1526.76 &  0 & 1548.56 & 0 & 1524.46 & 0 & $-5.14$ & 0\\
    \midrule
    Base~3 & $69.58^{+0.56}_{-0.54}$ & -- & $-19.377\pm 0.016$ & 1513.14 & 1521.14 & 0 & 1542.89 & 0 & 1521.01 & 0 & $-6.23$ & 0\\
    Base 3 + CC & $69.58^{+0.54}_{-0.55}$ & -- & $-19.378\pm 0.016$ & 1528.09 & 1536.09 & 0 & 1557.92 & 0 & 1535.94 & 0 &$-6.31$ & 0\\ 
    Base 3 + CC + R22 & $70.20^{+0.49}_{-0.48}$ & -- & $-19.360\pm 0.014$ & 1537.17 & 1545.17 & 0 & 1567.00 & 0 & 1545.04 & 0 & $-6.34$ & 0\\ 
    \midrule
   \multirow{3}{*}{Datasets}  & \multicolumn{10}{c}{$\Lambda$CDM + $\Omega_{\rm{K}}$  ($\tilde{\zeta}_0 = 0$)}\\  
    \cmidrule{2-13}
    & $H_0$ & $\Omega_{\rm{K0}}$ &  $M$  & $\chi^2_{\rm{min}}$ & $\rm{AIC}_{i}$ & $\Delta$AIC & $\rm{BIC}_{i}$& $\Delta$BIC & $\rm{DIC}_{i}$ & $\Delta$DIC & $\ln \mathcal{E}_{i}$ & $\ln \mathcal{B}_{0i}$\\ 
    & (\si{km.s^{-1}.Mpc^{-1}}) & & &&&&\\
    \midrule
    PPS & $73.4\pm 1.1$ & $0.147^{+0.077}_{-0.078}$ &  $-19.245^{+0.034}_{-0.033}$ & 1451.73 & 1461.73 & $-1.7$ & 1488.79 & $-7.11$ & 1458.45 & 0.36 & $-7.32$ &  2.83\\ 
    PPS + DESI & $73.5\pm 1.1$ & $0.052^{+0.033}_{-0.034}$  & $-19.251^{+0.031}_{-0.033}$ & 1463.99 & 1473.99 & $-0.31$ & 1501.09 & $ -5.73$ & 1473.00 & 0.60 & $-9.87$ & 4.61\\ 
    \midrule
    Base 1 & $66.7\pm 1.1$ & $0.120^{+0.068}_{-0.065}$ & $-19.457^{+0.030}_{-0.031}$& 1418.54 &  1428.54 & 6.82 & 1455.49 & 1.43 & 1427.47 & 5.57 & $-7.69$ & 3.07 \\ 
    Base 1 + CC  & $66.8\pm 1.1$ & $0.118^{+0.067}_{-0.066}$  & $-19.454^{+0.030}_{-0.029}$& 1434.05 & 1444.05 & 6.48 & 1471.10 & 1.07 & 1442.77 & 5.83 & $-7.72$ & 3.00\\ 
        Base 1 + CC + R22  & $69.78^{+0.84}_{-0.79}$ & $0.017^{+0.053}_{-0.055}$ & $-19.376^{+0.022}_{-0.021}$  & 1456.07 & 1466.07 & $-1.32$ & 1493.12 & $-6.73$ & 1463.49 & $-0.64$ & $-7.92$ & 2.93 \\ 
   \midrule
    Base 2  & $70.37^{+0.83}_{-0.86}$ & $0.004\pm 0.055$ &  $-19.359\pm 0.022$ & 1498.16 & 1508.16 & $-1.64$ & 1535.32 & $-7.08$ & 1505.59 & $-1.21$ & $-7.83$ & 2.89 \\ 
    Base 2 + CC & $70.28^{+0.82}_{-0.83}$ & $0.006^{+0.054}_{-0.055}$ & $-19.361\pm 0.021$ & 1513.65 & 1523.65 & $-1.68$ & 1550.90 & $-7.12$ & 1521.02 & $-1.19$ & $-7.91$ & 2.91\\ 
        Base 2 + CC + R22 & $71.22^{+0.68}_{-0.67}$ & $-0.027^{+0.051}_{-0.050}$ & $-19.338^{+0.018}_{-0.017}$ & 1518.73 & 1528.73 & $-1.98$  & 1555.99 & $-7.43$ & 1525.97 & $-1.51$ & $-8.01$ &  2.87 \\ 
    \midrule
    Base 3 & $69.60^{+0.67}_{-0.64}$ & $-0.002\pm 0.030$ & $-19.377\pm 0.018$& 1513.19 & 1523.19 & $-2.05$ & 1550.38 & $-7.49$ & 1522.53 & $-1.52$ & $-11.04$ & 4.81\\ 
    Base 3 + CC & $69.59\pm 0.64$ & $-0.001^{+0.029}_{-0.030}$ & $-19.378^{+0.018}_{-0.017}$ & 1528.12 & 1538.12 & $-2.03$ & 1565.41 & $-7.49$ & 1537.46 & $-1.52$ & $-11.14$ & 4.83 \\     
    Base 3 + CC + R22 & $70.39^{+0.57}_{-0.54}$ & $-0.021\pm 0.028$ & $-19.357^{+0.016}_{-0.015}$ & 1536.70 & 1546.70 & $-1.53$ & 1573.99 & $-6.99$ & 1545.99 & $-0.95$ & $-11.21$ & 4.87 \\ 
    \midrule
\midrule
\end{tabular}
\tablefoot{We shown information criteria $\Delta\mathrm{IC} = \mathrm{IC}_{\Lambda\mathrm{CDM}} - \mathrm{IC}_{\Lambda\mathrm{CDM}+\Omega_K}$, where $\Delta\mathrm{IC} > 2$ indicates preference for the curved model; and Bayesian evidence via $\ln\mathcal{B}_{0i} = \ln\mathcal{E}_{\Lambda\mathrm{CDM}} - \ln\mathcal{E}_{\Lambda\mathrm{CDM}+\Omega_K}$,  where $\ln\mathcal{B}_{0i} > 1$ favors $\Lambda$CDM.}
\end{sidewaystable*}

\section{Summary of cosmological constraints}
\label{appendix:summary_constraints}

The cosmological constraints on the parameters: $H_0$, $\Omega_{\rm{K0}}$, $\Omega_{\rm{b0}}h^2$, $\Omega_{\rm{vc0}}h^2$, $M$, $\tilde{\zeta}_0$, and $m$ are summarized in Tables \ref{table:results_m=0} and \ref{table:results_mfree}.

\begin{sidewaystable*}[h!]
\caption{Summary of best-fit paramaters values at $68\%$ CL for both  $\Lambda$vCDM and $\Lambda$vCDM + $\Omega_{\rm{K}}$ scenarios with constant bulk viscosity ($m=0$), using different datasets.}
\label{table:results_m=0}
\centering
 \begin{tabular}{lccccccccc}
\midrule\midrule
        \multirow{2}{*}{Dataset}  & $H_0$ & $\Omega_{\rm{K}0}$ &  $\Omega_{\rm{b0}}h^2$ & $\Omega_{\rm{vc}0}h^2$  & $M$ &  $\tilde{\zeta}_0$ & $\zeta_0$ ($\times 10^{6}$) & $\chi^2_{\rm{min},i}$ & $\Delta\chi^2_{\rm{min}}$   \\  
        & (\si{km.s^{-1}.Mpc^{-1}}) & & &&& & (\si{Pa.s}) &&\\ 
        \midrule
        &\multicolumn{9}{c}{\multirow{1}{*}{$\Lambda$vCDM ($m=0$)}}\\ 
            \cmidrule{2-10}
                PPS & $73.1\pm 1.1$ & --  & $0.037 \pm 0.010$  & $0.135 \pm 0.009 $  &  $-19.265\pm 0.032$  & $0.017^{+0.032}_{-0.017}$ & $0.72^{+1.35}_{-0.72}$ & 1451.96 & 0.07   \\
        PPS + DESI & $73.3\pm 1.1$ & -- & $0.028\pm 0.002$ & $0.137^{+0.007}_{-0.006}$  & $-19.262\pm 0.032$ &  $0.011^{+0.021}_{-0.011}$  & $0.47^{+0.89}_{-0.47}$ & 1464.63 & 1.05 \\
        \midrule
        Base 1  & $67.94^{+0.97}_{-0.92}$ & -- & $0.030\pm 0.010$ &  $0.127 \pm 0.010$ &  $-19.426^{+0.026}_{-0.027}$  & $0.092^{+0.075}_{-0.066}$ & $3.62^{+2.95}_{-2.60}$ & 1420.81 & 6.55    \\ 
        Base 1 + CC   & $68.0\pm 0.9$ & -- & $0.029^{+0.010}_{-0.009}$ &  $0.127 \pm 0.010$ & $-19.425^{+0.025}_{-0.026}$ & $0.088^{+0.075}_{-0.063}$ & $3.46^{+2.95}_{-2.48}$ & 1435.77 & 6.76  \\ 
                Base 1 + CC + R22 & $69.92^{+0.71}_{-0.74}$ & -- & $0.027\pm 0.010$ &  $0.125\pm 0.010$ & $-19.373^{+0.019}_{-0.020}$ & $0.082^{+0.073}_{-0.061}$  & $3.32^{+2.95}_{-2.47}$ & 1453.00 & 3.75 \\ 
                \midrule
        Base 2 & $70.39^{+0.75}_{-0.73}$ & --  & $0.027\pm 0.010$ & $0.125\pm 0.010$ &   $-19.357\pm 0.020$  & $0.088^{+0.072}_{-0.063}$ & $3.59^{+2.93}_{-2.57}$  & 1494.40 & 4.12 \\
        Base 2 + CC & $70.30^{+0.73}_{-0.72}$ & --  & $0.027\pm 0.010$  & $0.125 \pm 0.010 $ &  $-19.360^{+0.020}_{-0.019}$ & $0.085^{+0.070}_{-0.061}$ &  $3.46^{+2.85}_{-2.48}$ & 1510.30 & 3.68  \\ 
                Base 2 + CC + R22 & $71.05^{+0.62}_{-0.60}$ & -- & $0.026\pm 0.010$ & $0.124\pm 0.010$ & $-19.340\pm 0.016$ & $0.081^{+0.071}_{-0.058}$ & $3.33^{+2.92}_{-2.38}$  & 1516.14 & 2.62 \\ 
        \midrule
        Base~3 & $69.53\pm 0.56$ & -- &  $0.021 \pm 0.002$ & $0.124^{+0.009}_{-0.007} $  & $-19.379^{+0.016}_{-0.017}$ & $0.030^{+0.038}_{-0.026}$ & $1.21^{+1.53}_{-1.05}$ & 1511.96 & 1.18\\
        Base~3 + CC & $69.52^{+0.54}_{-0.57}$ & -- & $0.021 \pm 0.002$ & $0.124^{+0.009}_{-0.007}$ & $-19.379 \pm 0.016$ & $0.030^{+0.036}_{-0.026}$ & $1.21^{+1.45}_{-1.05}$ & 1526.91 & 1.18 \\
        Base~3 + CC + R22  & $70.17^{+0.49}_{-0.51}$ & -- & $0.022 \pm 0.002$ & $0.122^{+0.009}_{-0.007}$ & $-19.361^{+0.014}_{-0.015}$ & $0.026^{+0.035}_{-0.023}$ & $1.06^{+1.42}_{-0.93}$ & 1537.12 & 0.05 \\
        \midrule
        &\multicolumn{9}{c}{$\Lambda$vCDM + $\Omega_{\rm{K}}$ ($m=0$)} \\ 
        \cmidrule{2-10}
                PPS & $73.3 \pm 1.2$ & $0.196^{+0.092}_{-0.090}$ & $0.026\pm 0.011$ & $0.124\pm 0.011$ & $-19.247^{+0.033}_{-0.034}$ & $0.048^{+0.084}_{-0.047}$ & $2.04^{+3.56}_{-1.99}$  & 1451.72  & 0.31 \\
        PPS + DESI & $73.3\pm 1.1$ & $0.062^{+0.037}_{-0.035}$ & $0.031\pm 0.003 $ & $0.129 \pm 0.009$ & $-19.255^{+0.032}_{-0.033}$ &  $0.019^{+0.029}_{-0.019}$ & $0.81^{+1.23}_{-0.81}$ & 1463.20 &  2.48  \\
        \midrule
        Base 1 & $66.1^{+1.2}_{-1.1}$ & $0.181^{+0.073}_{-0.074}$ & $0.023\pm 0.011$ & $0.121\pm 0.011$ & $-19.470^{+0.032}_{-0.030}$ & $0.195^{+0.094}_{-0.095}$ & $7.46^{+3.60}_{-3.64}$   & 1417.36 & 10.00  \\ 
        Base 1 + CC  & $66.3\pm 1.1$ &  $0.174^{+0.074}_{-0.070}$ & $0.023\pm 0.011$  & $0.121\pm 0.011$ &$-19.466^{+0.030}_{-0.029}$ & $0.188^{+0.091}_{-0.094}$ & $7.20^{+3.49}_{-3.60}$  & 1433.17 & 9.36  \\ 
                Base 1 + CC + R22 & $69.56^{+0.81}_{-0.83}$ &  $0.049^{+0.059}_{-0.058}$  & $0.025\pm 0.011$  & $0.123 \pm 0.011$ & $-19.381^{+0.021}_{-0.022}$ & $0.109^{+0.081}_{-0.074}$ & $4.39^{+3.26}_{-2.98}$ & 1453.65 & 3.10   \\ 
                \midrule
        Base 2 & $70.09^{+0.87}_{-0.90}$ &  $0.040\pm 0.061$ & $0.025\pm 0.011$  & $0.124\pm 0.011$ & $-19.365^{+0.023}_{-0.022}$ & $0.111^{+0.082}_{-0.075}$ & $4.50^{+3.33}_{-3.04}$ & 1494.81 & 3.71  \\ 
        Base 2 + CC & $70.00^{+0.84}_{-0.85}$ &  $0.025\pm 0.011$ & $0.025\pm 0.011$  & $0.123\pm 0.011$ & $-19.366^{+0.021}_{-0.022}$ & $0.105^{+0.080}_{-0.072}$ & $4.25^{+3.24}_{-2.91}$ & 1510.80 & 3.18  \\ 
                Base 2 + CC + R22 & $71.06\pm 0.70$ &  $-0.002^{+0.056}_{-0.053}$  & $0.026\pm 0.011$ & $0.125\pm 0.011$ & $-19.341\pm 0.018$ & $0.078^{+0.077}_{-0.059}$ & $3.16^{+3.17}_{-2.43}$ & 1515.16 & 3.60  \\ 
        \midrule
        Base~3 & $69.42^{+0.68}_{-0.67}$ & $0.009\pm 0.032$ & $ 0.021 \pm 0.002$ & $0.124 \pm 0.009$ & $-19.381^{+0.018}_{-0.019}$ & $0.033^{+0.040}_{-0.028}$ & $1.33^{+1.61}_{-1.12}$ & 1512.06 & 1.08\\
        Base~3 + CC & $69.38^{+0.67}_{-0.65}$ & $0.009^{+0.032}_{-0.031}$ & $0.021 \pm 0.002$ & $0.124 \pm 0.009$  & $-19.382\pm 0.018$ &  $0.032^{+0.040}_{-0.028}$ & $1.29^{+1.61}_{-1.12}$ & 1526.91 &  1.18 \\ 
        Base~3 + CC + R22 & $70.28^{+0.60}_{-0.58}$ & $-0.014^{+0.030}_{-0.029}$  & $0.0219^{+0.0018}_{-0.0019}$  & $0.126^{+0.009}_{-0.008}$  &  $-19.358\pm 0.016$  &  $0.024^{+0.034}_{-0.024}$ & $0.98^{+1.38}_{-0.98}$ & 1536.15 &  1.02 \\
        \midrule
\end{tabular}
\tablefoot{All $\Delta$ are computed relative to the standard  $\Lambda$CDM model for the same combination of data sets.  Here, $\Delta \chi_{\rm{min}}^2 \equiv \chi_{\rm{min, \Lambda CDM}} - \chi_{\rm{min},i}$, where the index $i$ denotes each proposed bulk viscous model. Positive values of $\Delta \chi_{\rm{min}}^2$ indicate that the bulk viscous model provides a better fit to the data. The viscosity parameter $\zeta_0$ is calculated using Eq. (\ref{eq:zeta0_tilde}).}
\end{sidewaystable*}

\begin{sidewaystable*}
\caption{Summary of best-fit paramaters values at $68\%$ CL for both  $\Lambda$vCDM and $\Lambda$vCDM + $\Omega_{\rm{K}}$ scenarios with free $m$, using different datasets.}
\label{table:results_mfree}
\centering
\begin{tabular}{lccccccccccc}
\midrule\midrule
    \multirow{2}{*}{Dataset}  & $H_0$ & $\Omega_{\rm{K0}}$ & $\Omega_{\rm{b0}}h^2$ & $\Omega_{\rm{vc}0}h^2$ & $M$ & $m$  &  $\tilde{\zeta}_0$ & $\zeta_0$ ($\times 10^{6}$) & $\chi^2_{\rm{min}}$ & $\Delta\chi^2_{\rm{min}}$  \\  
    & (\si{km.s^{-1}.Mpc^{-1}}) & &&& & && (\si{Pa.s}) &&\\ \midrule
    &\multicolumn{10}{c}{\multirow{1}{*}{$\Lambda$vCDM  ($m$ free)}}\\ 
         \cmidrule{2-11}
        PPS & $73.1\pm 1.1$ & -- & $0.036^{+0.010}_{-0.009}$ & $ 0.135^{+0.009}_{-0.010} $ & $-19.264^{+0.032}_{-0.034}$  & $-1.11^{+1.39}_{-0.82}$ & $0.020^{+0.040}_{-0.019}$  & $0.85^{+1.69}_{-0.80}$ & 1451.78 & 0.25 \\   
    PPS + DESI & $73.4\pm 1.1$ & -- & $0.028 \pm 0.002$ & $0.136 \pm 0.006$  & $-19.261\pm 0.031$ & $-1.25^{+1.38}_{-0.63}$ & $0.012^{+0.026}_{-0.011}$ & $0.51^{+1.10}_{-0.47}$ & 1463.19 & 2.49 \\
    \midrule
    Base 1 & $67.3^{+1.1}_{-1.0}$ & -- & $0.022\pm 0.011$ & $0.127^{+0.010}_{-0.010}$ &  $-19.439\pm 0.028$ &  $0.89^{+0.61}_{-0.60}$  & $0.032^{+0.050}_{-0.030}$ & $1.25^{+1.95}_{-1.17}$ & 1416.83  & 10.53  \\
    Base 1 + CC   & $67.4\pm 1.0$  & --  &  $0.030\pm 0.010$ & $0.127 \pm 0.010$ & $-19.437^{+0.027}_{-0.028}$ &  $0.91^{+0.63}_{-0.59}$  &  $0.030^{+0.048}_{-0.029}$ & $1.17^{+1.87}_{-1.13}$ &  1431.90 & 10.63   \\ 
        Base 1 + CC + R22  & $69.76^{+0.78}_{-0.79}$ & -- & $0.025 \pm 0.010$ & $0.123 \pm 0.010$ &  $-19.376\pm 0.021$  &  $0.51^{+0.86}_{-1.03}$ &  $0.040^{+0.064}_{-0.039}$ & $1.61^{+2.58}_{-1.57}$  &  1451.68 & 5.07 \\ 
        \midrule
    Base 2 & $70.19^{+0.82}_{-0.80}$  & -- &  $0.025^{+0.011}_{-0.010}$ & $0.124\pm 0.010$ &  $-19.361\pm 0.021$ & $0.50^{+0.83}_{-0.93}$ & $0.043^{+0.064}_{-0.042}$ & $1.75^{+2.60}_{-1.71}$ & 1493.43 & 5.09  \\
    Base 2 + CC   &  $70.13^{+0.82}_{-0.80}$ & -- & $0.024\pm 0.010$ & $0.122\pm 0.010$ & $-19.363\pm 0.021$  & $0.51^{+0.85}_{-0.98}$ & $0.042^{+0.061}_{-0.041}$ & $1.70^{+2.48}_{-1.66}$ & 1509.09 & 4.89  \\ 
        Base 2 + CC + R22 & $71.00^{+0.67}_{-0.66}$  & -- & $0.024 \pm 0.010 $ & $0.122^{+0.010}_{-0.009}$ &  $-19.342\pm 0.017$  & $0.23^{+0.83}_{-1.18}$ &  $0.048^{+0.070}_{-0.047}$ & $1.97^{+2.88}_{-1.93}$ &   1515.76 & 3.00  \\ 
        \midrule
    Base~3 & $69.61^{+0.63}_{-0.62}$ & -- & $0.022 \pm 0.002$ & $0.122^{+0.008}_{-0.006}$ & $-19.378^{+0.018}_{-0.017}$ & $-0.87^{+1.15}_{-0.84}$ & $0.029^{+0.050}_{-0.028}$  & $1.17^{+2.01}_{-1.12}$ & 1511.89 & 1.25\\
    Base~3 + CC  & $69.59^{+0.61}_{-0.60}$ & -- & $ 0.022^{+0.001}_{-0.002}$ & $0.122^{+0.008}_{-0.006}$ & $-19.378\pm 0.017$  & $-0.92^{+1.19}_{-0.80}$ & $0.030^{+0.050}_{-0.029}$ & $1.21^{+2.01}_{-1.17}$ & 1527.00 & 1.09 \\
    Base~3 + CC + R22 & $70.28^{+0.54}_{-0.52}$ & -- & $0.023 \pm 0.002$ & $0.121^{+0.008}_{-0.006}$ & $-19.359\pm 0.015$ &  $-1.31^{+1.03}_{-0.62}$ & $0.038^{+0.063}_{-0.037}$ & $1.55^{+2.56}_{-1.50}$ & 1537.18 & $-0.01$ \\
    \midrule
    &\multicolumn{10}{c}{\multirow{1}{*}{$\Lambda$vCDM + $\Omega_{\rm{K}}$ ($m$ free)}} \\ 
    \cmidrule{2-11}
        PPS & $73.3\pm 1.2$ & $0.205^{+0.103}_{-0.094}$ &  $0.026\pm 0.011$ & $0.124\pm 0.011$ & $-19.247^{+0.033}_{-0.034}$ & $-1.27^{+1.10}_{-0.72}$  & $0.073^{+0.137}_{-0.071}$  & $3.10^{+5.81}_{-3.01}$ & 1451.76 & 0.27 \\
    PPS + DESI & $73.4\pm 1.1$ & $0.062^{+0.040}_{-0.037}$ & $0.031 \pm 0.004$ & $0.129\pm 0.011$ & $-19.255^{+0.040}_{-0.039}$ & $-1.19^{+1.14}_{-0.70}$  & $0.032^{+0.065}_{-0.029}$ & $1.36^{+2.72}_{-1.23}$ & 1462.96 & 2.72  \\
    \midrule
    Base 1 & $66.1\pm 1.3$ & $0.21^{+0.13}_{-0.12}$  & $0.021^{+0.018}_{-0.017}$  & $0.119^{+0.016}_{-0.015}$ & $-19.474^{+0.037}_{-0.036}$ & $-0.41^{+0.95}_{-1.04}$ & $0.210^{+0.230}_{-0.180}$ & $8.03^{+8.80}_{-6.89}$ & 1416.83 & 10.53  \\ 
    Base 1 + CC   & $66.2\pm 1.2$ & $0.20^{+0.12}_{-0.11}$   & $0.021\pm 0.012$  & $0.119\pm 0.012$ & $-19.470^{+0.034}_{-0.033}$& $-0.45^{+0.95}_{-1.03}$ & $0.180^{+0.230}_{-0.160}$  & $6.90^{+8.81}_{-6.13}$  & 1431.98 & 10.55\\
        Base 1 + CC + R22 & $69.58^{+0.90}_{-0.86}$  & $0.026^{+0.087}_{-0.082}$ & $0.026^{+0.087}_{-0.082}$ &  $0.122^{+0.012}_{-0.011}$ & $-19.380^{+0.023}_{-0.024}$ & $0.29^{+0.92}_{-1.31}$ & $0.059^{+0.110}_{-0.057}$ & $2.38^{+4.43}_{-2.30}$ & 1448.63 & 8.12  \\ 
        \midrule 
    Base 2 & $70.14^{+0.93}_{-0.91}$ & $0.007^{+0.079}_{-0.077}$ & $0.026\pm 0.012$  & $0.124\pm 0.012$ & $-19.363^{+0.024}_{-0.023}$  & $0.47^{+0.91}_{-1.06}$  & $0.049^{+0.089}_{-0.047}$ & $1.99^{+3.61}_{-1.91}$ & 1489.50 & 9.02  \\ 
    Base 2 + CC & $70.07^{+0.88}_{-0.90}$ & $0.008^{+0.078}_{-0.081}$ & $0.026\pm 0.012$  & $0.124\pm 0.012$ & $-19.365 \pm 0.023$ & $0.49^{+0.92}_{-1.14}$ &  $0.047^{+0.090}_{-0.045}$ & $1.91^{+3.65}_{-1.82}$ & 1504.71 & 9.27  \\ 
        Base 2 + CC + R22 & $71.07^{+0.74}_{-0.73}$ & $-0.026^{+0.072}_{-0.074}$ & $0.027\pm 0.012$  & $0.125\pm 0.012$ &   $-19.340\pm 0.019$ & $0.51^{+0.94}_{-1.21}$ & $0.041^{+0.073}_{-0.039}$ & $1.69^{+3.00}_{-1.60}$ & 1508.26 & 10.50  \\ 
    \midrule 
    Base~3 & $69.53^{+0.69}_{-0.72}$ & $0.008^{+0.037}_{-0.035}$ & $0.022 \pm 0.002$ & $ 0.122 \pm 0.009$ & $-19.380\pm 0.019$  & $-0.99^{+1.16}_{-0.80}$ & $0.036^{+0.063}_{-0.034}$ & $1.45^{+2.34}_{-1.37}$ & 1511.75 & 1.39\\ 
    Base~3 + CC & $69.49^{+0.68}_{-0.69}$ & $0.010^{+0.036}_{-0.035}$ & $0.022\pm 0.002$ & $0.122^{+0.009}_{-0.008}$ & $-19.380\pm 0.019$ & $-1.01^{+1.14}_{-0.78}$  & $0.037^{+0.065}_{-0.035}$ & $1.49^{+2.61}_{-1.41}$ & 1526.52  & 1.57\\ 
    Base~3 + CC + R22 & $70.36^{+0.58}_{-0.57}$ & $-0.009\pm 0.032$ & $0.022 \pm 0.002$ & $0.124 \pm 0.008$  & $-19.358\pm 0.016$ & $-1.31^{+1.09}_{-0.62}$ & $0.032^{+0.067}_{-0.030}$  & $1.30^{+2.73}_{-1.22}$ & 1535.44 & 1.73 \\
    \midrule    
    \end{tabular}
\tablefoot{All $\Delta$ are computed relative to the standard  $\Lambda$CDM model for the same combination of data sets.  Here, $\Delta \chi_{\rm{min}}^2 \equiv \chi_{\rm{min, \Lambda CDM}} - \chi_{\rm{min},i}$, where the index $i$ denotes each proposed bulk viscous model. Positive values of $\Delta \chi_{\rm{min}}^2$ indicate that the bulk viscous model provides a better fit to the data.  We calculate $\zeta_0$  using Eq. (\ref{eq:zeta0_tilde}).}
\end{sidewaystable*}

\FloatBarrier
\onecolumn

\section{Model comparison summary}
\label{Model_comparison_summary}

\begin{table}[h!]
\caption{Comparison of viscous cosmological models with standard $\Lambda$CDM.}
\label{table:summary_IC_BayesFactor}
\centering 
    \begin{tabular}{lcccccccc}
   \midrule\midrule
    \multirow{1}{*}{Datasets}  &  $\rm{AIC}_{i}$ & $\Delta$AIC  & $\rm{BIC}_{i}$ & $\Delta$BIC & $\rm{DIC}_{i}$ & $\Delta$DIC & $\ln \mathcal{E}_{i}$ &  $\ln \mathcal{B}_{0i}$ \\ 
    \midrule
    & \multicolumn{8}{c}{$\Lambda$vCDM ($m=0$)} \\  \cmidrule{2-9}
        PPS & 1461.96 & $-1.93$ & 1489.02 & $-7.34$ & 1459.41 &  $-0.6$ & $-6.28$ & 1.79 \\
        \midrule
        Base 1   & 1430.81 & 4.55  & 1457.76 & $-0.84$ & 1430.52 & 2.52 & $-5.95$ & 1.33 \\ 
        Base 1  + CC  & 1445.77 & 4.76   &  1472.82 & $-0.65$   & 1445.29 & 3.31 &  $-5.98$ & 1.26 \\
        Base 1 + CC + R22   & 1463.00  & 1.75  &  1490.05 & $-3.66$  & 1461.14 & 1.71  &  $-6.17$ & 1.18 \\
        \midrule
        Base 2  & 1506.52  &  2.12 & 1531.55  & $-3.31$  & 1502.40 & 1.98  &  $-6.34$ & 1.40 \\
        Base 2 + CC  &   1520.30 & 1.68   & 1547.55 & $-3.77$  & 1517.84 & 1.99  &  $-6.31$ & 1.31  \\ 
        Base 2 + CC + R22 &  1526.14 &  0.62  &  1553.39 & $-4.83$  & 1522.84 & 1.62  &  $-6.51$ & 1.37  \\ 
        \midrule
        Base 3 & 1521.96 & $-0.82$ & 1549.15 & $-6.26$ & 1520.51 & 0.50 & $-8.29$ & 2.06 \\
        Base 3 + CC & 1536.91 & $-0.82$ & 1564.20 & $-6.28$ & 1535.48 & 0.46 & $-8.26$ & 1.95\\
        Base 3 + CC + R22 & 1547.12 & $-1.95$ & 1574.42 &  $-7.42$ & 1545.06 & $-0.02$ & $-8.55$ & 2.21\\
        \midrule
        &\multicolumn{8}{c}{$\Lambda$vCDM + $\Omega_{\rm{K}}$  ($m=0$) } \\  \cmidrule{2-9}
        PPS & 1463.72 & $-3.69$ & 1496.20 & $-14.52$ & 1459.62 & $-0.81$ & $-8.77$ & 4.28 \\
        \midrule
        Base 1    &  1429.36 & 6.00  & 1461.70 &  $-4.78$  & 1425.68 & 7.36 & $-8.87$ & 4.25 \\ 
        Base 1  + CC    &  1445.17 &  5.36  & 1477.63 &  $-5.46$ & 1441.32 & 7.28 & $-8.93$ & 4.21  \\ 
        Base 1 + CC + R22  &  1465.65 &  $-0.90$  & 1498.12 & $-11.73$  & 1462.22 & 0.63  & $-9.27$ & 4.28 \\ 
        \midrule
        Base 2  &  1506.81 & $-0.29$  & 1539.39 &  $-11.15$  & 1503.93 & 0.45  & $-9.29$ & 4.35 \\ 
        Base 2 + CC  & 1522.80 &  $-0.82$  & 1555.50 & $-11.72$  & 1519.07 & 0.76  & $-9.20$ & 4.20 \\ 
        Base 2 + CC + R22  &  1527.16 &  $-0.40$  & 1559.87 &  $-11.31$  & 1524.37 & 0.09  &  $-9.54$ & 4.40 \\ 
        \midrule
        Base 3 & 1524.06 & $-2.92$ & 1556.69 & $-13.80$ & 1522.36 & $-1.35$ & $-13.34$ & 7.11\\
        Base 3 + CC & 1538.91 & $-2.82$ & 1571.66 & $-13.74$ & 1537.29 & $-1.35$ & $-13.36$ & 7.05\\
        Base 3 + CC + R22 & 1548.15 & $-2.98$ & 1580.90 & $-13.90$ & 1546.22 & $-1.18$ & $-13.52$ & 7.18\\
        \midrule
        &\multicolumn{8}{c}{$\Lambda$vCDM  ($m$ free)}\\  \cmidrule{2-9}
        PPS & 1463.78 & $-3.75$ & 1496.26 & $-14.58$ & 1459.43 & $-0.62$ & $-5.01$ & 0.52 \\
        \midrule
        Base 1 & 1428.83 & 6.53 & 1461.17  & $-4.25$ & 1426.09 & 6.95 & $-5.60$  & 0.98  \\
        Base 1 + CC & 1443.90 & 6.63 & 1476.36 & $-4.19$ & 1440.63 & 7.97 &  $-5.57$ & 0.85 \\
        Base 1 + CC + R22 & 1463.68 & 1.07 & 1496.15 & $-9.76$ & 1459.87 & 2.98 & $-5.48$ & 0.49 \\
        \midrule
        Base 2   & 1505.43 & 1.09 & 1538.01 & $-9.77$  & 1502.09 &  2.29 & $-5.52$ & 0.58  \\
        Base 2 + CC  & 1521.09 & 0.89  & 1553.79 & $-10.01$  & 1517.09 & 2.74  & $-5.60$ & 0.60 \\
        Base 2 + CC + R22  & 1527.76 & $-1.00$  & 1560.46 & $-11.90$  &1523.25 & 1.21 & $-5.53$ & 0.39 \\
        \midrule
        Base 3 & 1523.88 & $-2.74$ & 1556.51 & $-13.62$ & 1521.07 & $-0.06$ & $-6.94$ & 0.71\\
        Base 3 + CC & 1539.00 & $-2.91$ & 1571.75 & $-13.83$ & 1535.96 & $-1.85$ & $-7.16$ & 0.85\\
        Base 3 + CC + R22 & 1549.17 & $-4.00$ & 1581.93 & $-14.93$ & 1545.36 & $-0.32$ & $-7.07$ & 0.73\\
        \midrule
        &\multicolumn{8}{c}{$\Lambda$vCDM + $\Omega_{\rm{K}}$ ($m$ free)} \\  \cmidrule{2-9}
        PPS & 1465.76 & $-5.73$ & 1503.65 & $-21.97$ & 1459.41 & $-0.60$ & $-7.49$ & 3.00 \\
        \midrule
        Base 1  & 1430.83 &  4.53 & 1468.56 & $-11.64$  & 1425.35 & 7.69 &  $-7.74$ & 3.12 \\
        Base 1 + CC  & 1445.98  & 4.55 & 1483.86 & $-11.69$  & 1441.21 & 7.39 & $-7.72$ & 3.00 \\
        Base 1 + CC + R22  & 1462.63  & 2.12 & 1500.51 & $-14.12$ & 1460.85 & 2.00  & $-8.13$ & 3.14 \\
        \midrule
        Base 2   & 1503.50  & 3.02  & 1541.52 & $-13.28$ & 1488.85 & 15.53 & $-8.23$ & 3.29  \\
        Base 2 + CC  & 1518.71 & 3.27  & 1556.86 & $-13.08$ & 1518.12 & 1.71 & $-8.21$ & 3.21 \\
        Base 2 + CC + R22  & 1522.26 &  4.50  & 1560.42  & $-11.86$  & 1526.61 & $-2.15$ & $-8.31$ & 3.17 \\
        \midrule
        Base 3 & 1525.75 & $-4.61$ & 1563.82 & $-20.93$ & 1522.81 & $-1.80$ & $-11.78$ & 5.55\\
        Base 3 + CC & 1540.52 & $-4.43$ & 1578.72 & $-20.80$ & 1537.79 & $-1.85$ & $-11.94$ & 5.63\\
        Base 3 + CC + R22 & 1549.44 & $-4.27$ &  1587.64 & $-20.64$ & 1546.38 & $-1.34$ & $-11.89$ & 5.55\\
        \midrule
    \end{tabular}
\tablefoot{We use information criteria (DIC, BIC, and AIC), Bayesian evidence ($\ln\mathcal{E}$), and Bayes factor ($\ln\mathcal{B}_{0i}$). Here, $\Delta\mathrm{IC} = \mathrm{IC}_{\Lambda\mathrm{CDM}} - \mathrm{IC}_{i}$ and $\ln\mathcal{B}_{0i} = \ln \mathcal{E}_{\Lambda\rm{CDM}} - \ln \mathcal{E}_{i}$, where the index $i$ denotes each proposed bulk viscous model. Values of $\Delta\mathrm{IC} > 2$ indicate a preference for bulk viscous model, while $\ln\mathcal{B}_{0i} > 1$ favors the $\Lambda\rm{CDM}$ model.}
\end{table}

\FloatBarrier
\onecolumn

\section{Constraints on $H_0$ and $\Omega_{\rm{K}}$ for both  bulk viscous and standard models}
\label{Appendix:Constraints}

\begin{figure}[h!]
    \centering
    \includegraphics[width=0.83\textwidth]{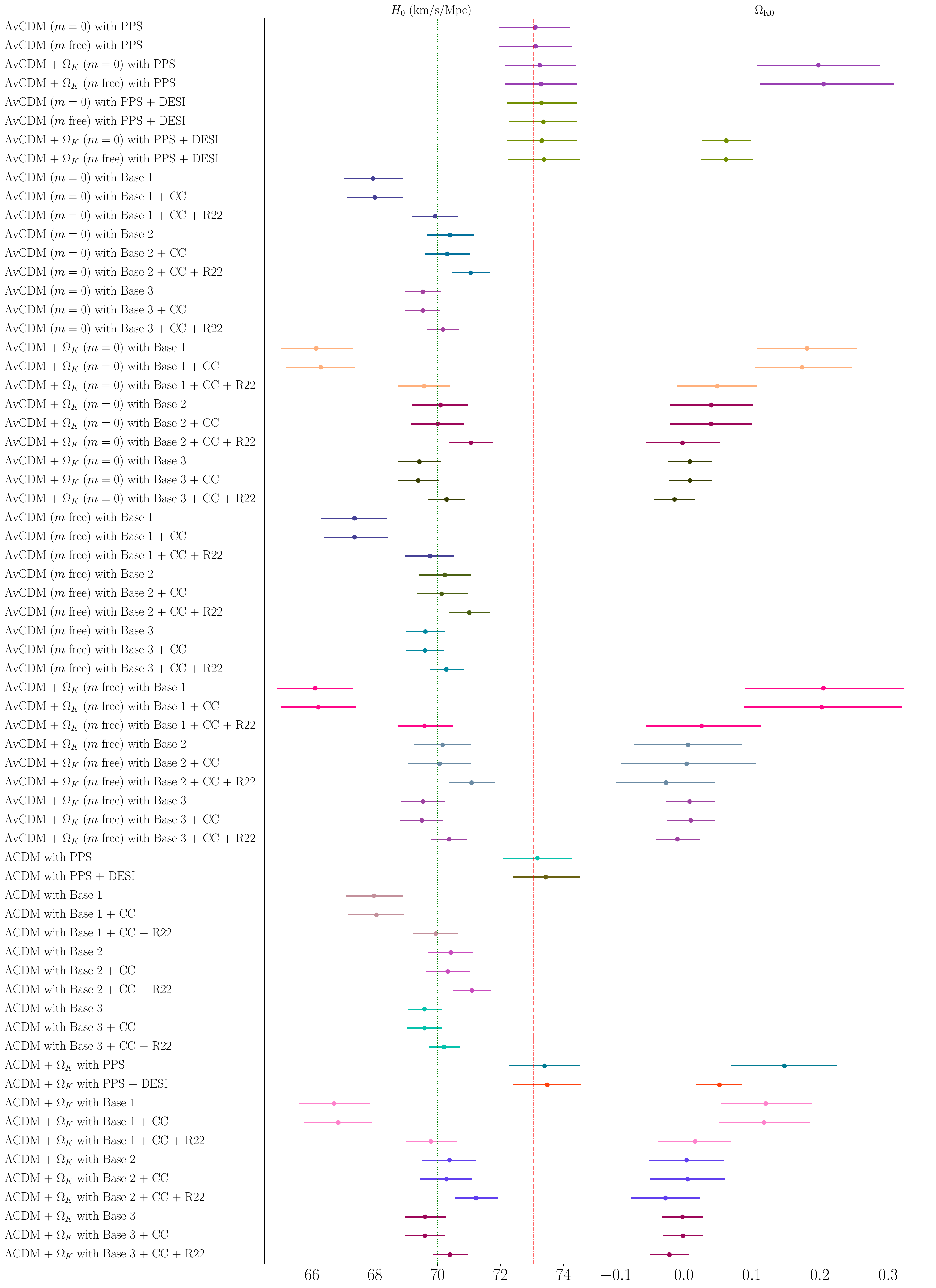}
    \caption{Whisker plot displaying the $68\%$ marginalized confidence intervals for the Hubble constant $H_0$ (\si{km.s^{-1}.Mpc^{-1}}) and  spatial curvature parameter $\Omega_{\rm{K}}$, derived from both viscous cold dark matter models with a cosmological constant ($\Lambda$vCDM and $\Lambda$vCDM+$\Omega_{\rm{K}}$ with $m=0$ and free $m$) and the standard $\Lambda$CDM model, including its curved extension. The analysis includes different datasets: DESI, CC, BAO, PP and PPS supernova samples. The red vertical dashed line indicates the $H_0$ measurement from R22 \citep{2022ApJ...934L...7R}), while the blue vertical dashed line marks the $\Lambda$CDM prediction for a flat universe ($\Omega_{\rm{K}} = 0$).}
    \label{fig:enter-label}
\end{figure}

\end{appendix}

\end{document}